\documentclass[twoside,12pt]{article}
\usepackage{epsfig}

\newcommand{\be}{\begin{equation}}
\newcommand{\ee}{\end{equation}}
\newcommand{\bea}{\begin{eqnarray}}
\newcommand{\eea}{\end{eqnarray}}

\begin{document}

\title{
Flavor Asymmetry of Light Quarks in the Nucleon Sea}

\author{Gerald T. Garvey and Jen-Chieh Peng
\\
Physics Division\\ Los Alamos National Laboratory\\
Los Alamos, New Mexico, 87545 USA}

\maketitle

\begin{abstract} A surprisingly large asymmetry between the up and down 
sea quark distributions in the nucleon has been observed in recent
deep inelastic scattering and Drell-Yan 
experiments.
This result strongly suggests that the mesonic
degrees of freedom play an important role in the description of the parton
distributions of the hadronic sea. 
In this article, we review the current status of
our knowledge of the flavor structure of the nucleon sea. The implications of
various theoretical models as well as possible future measurements are 
also discussed.
\end{abstract}
%\eject
%\tableofcontents
\section{Introduction}
The first direct evidence for point-like constituents in the nucleons came 
from the observation of scaling phenomenon in Deep-Inelastic Scattering
(DIS) experiments~\cite{bloom69,breiden69} at SLAC. 
These point-like charged constituents, called 
partons, were found to be spin-1/2 fermions. These partons were initially
identified as the quarks in the Constituent
Quark Models (CQM). However, it was soon realized that valence 
quarks alone could not account for the large enhancement of the cross
sections at small Bjorken-$x$, the fraction of nucleon momentum carried
by the partons. This swarm of low-momentum partons,
dubbed wee-partons by Feynman~\cite{feyn72}, was 
interpreted as the quark and antiquark
sea of the nucleon. DIS experiments therefore provided the first evidence
for the existence of antiquarks in the nucleon.

The observation of partons in DIS experiments paved the road to the 
formulation of Quantum Chromodynamics (QCD) as the theory for strong 
interactions. Nevertheless, the exact form of the parton distribution 
functions (PDF) can not be deduced from perturbative QCD. 
Like many other static 
properties of hadrons, the parton distribution functions belong to the 
domain of non-perturbative QCD. In spite of great progress~\cite{liu99} made 
in Lattice Gauge Theory (LGT) in treating the bound-state properties of 
hadrons, it remains a challenge to predict the parton distributions 
using LGT. 

Until parton distributions can be readily calculated from first principles, 
they are best determined from experiments. Electroweak processes such
as DIS and lepton-pair production provide the cleanest means to extract 
information on the parton distributions. There are at least two reasons
why it is important to measure the parton distribution functions. First,
the description of hard processes in high energy interactions requires parton 
distribution functions as an essential input. Second, many aspects 
of the parton distributions, such as sum rules, scaling-violation,
asymptotic behaviors at large and small $x$, and flavor and spin 
structures, can be compared with the predictions of perturbative as well
as non-perturbative QCD.

In this article, we review the status of our current knowledge of the
flavor dependence of the sea quark distributions in hadrons. 
The recent observation of a striking flavor asymmetry
of the nucleon sea has profound implications
on the importance of meson degrees of freedom for the description of parton
substructures in the nucleon. In Section 2, we review the early studies
of the nucleon sea in DIS and lepton-pair production. The crucial recent 
experiments establishing the up/down flavor asymmetry of the nucleon sea
are discussed in Section 3. The various theoretical models for 
explaining the flavor asymmetry of the nucleon sea are described in Section 4.
The implications of these models on other aspects of the
parton structure functions are discussed in Section 5. Finally, we
present future prospects in Section 6, followed by the 
conclusion in Section 7.

\section{Early Studies of the Nucleon Sea}
\subsection{\it Deep Inelastic Scattering}
Although scaling behavior in inelastic electron scattering
was predicted by Bjorken~\cite{bj69}, based on the framework of
current algebra, its confirmation by the SLAC experiments still came as a
major surprise. A simple and intuitive picture for explaining the
scaling behavior is the parton model advanced by 
Feynman~\cite{feyn72,feyn69}. In this model,
the electron-nucleon deep-inelastic scattering is described 
as an incoherent sum of elastic electron-parton scattering. However,
the nature of the partons within the nucleon was not specified by
Feynman. Some authors~\cite{drell69,drell70,cab70,lee72} speculated 
that the partons were the `bare nucleon'
plus the pion cloud, while others~\cite{bj69a,kuti71,land71} believed 
they were the quarks
introduced by Gell-Mann~\cite{gell64} and Zweig~\cite{zweig64}. 
The latter scenario was strongly
supported by the measurement of $R$, ratio of the longitudinally
over transversely polarized photon cross sections, showing the spin-1/2
character of the partons.

Evidence for quark-antiquark sea 
in the nucleon came from the observation that the structure function
$F_2(x)$ approaches a constant value as $x \to 0$~\cite{friedman}. 
If the proton is made 
up of only three quarks, or any finite number of quarks, $F_2(x)$ is expected
to vanish as $x \to 0$. Bjorken and Paschos~\cite{bj69a} 
therefore assumed that the nucleon consists of three quarks in a 
background of an infinite number of quark-antiquark pairs.
Kuti and Weisskopf~\cite{kuti71} further 
included gluons among the constituents of nucleons
in order to account for the missing momentum not carried by the
quarks and antiquarks alone.

The importance of the quark-antiquark pairs in the nucleon is in sharp 
contrast to the situation for the atomic system, where 
particle-antiparticle pairs play a relatively minor role (such as the 
polarization of the vacuum). In strong interactions, quark-antiquark pairs
are readily produced as a result of the
relatively large magnitude of the coupling constant $\alpha_s$,
and they form an integral part of the nucleon's structure.

Neutrino-induced DIS experiments allowed the separation of sea quarks
from the valence quarks. Recall that
\begin{eqnarray}
F_2^{\nu p}(x) = 2x \sum_i \left[q_i(x) + \bar q_i(x)\right], \hspace{0.8in} 
\nonumber \\ F_3^{\nu p}(x) = 2 \sum_i \left[q_i(x) - \bar q_i(x)\right] 
= 2 \sum_i q_i^v(x),
\label{eq:2.1.1}
\end{eqnarray}
where $i$ denotes the flavor of the quarks.
Note that the valence quark distribution is defined as the difference
of the quark and antiquark distributions, $q^v_i(x) = q_i(x) - \bar q_i(x)$.
Equation~\ref{eq:2.1.1} shows that the valence quark distribution is simply 
$F_3^{\nu p}(x)/2$, while the sea quark distribution is given by
$F_2^{\nu p}(x)/2x - F_3^{\nu p}(x)/2$. 
The $F_2^{\nu p}(x)$ and $F_3^{\nu p}(x)$ data 
from the CDHS experiment~\cite{cdhs79}
clearly showed that the valence 
quark distributions dominate at $x>0.2$, while the sea quarks are 
at small $x$. 

The earliest parton models assumed that the proton sea was flavor symmetric,
even though the valence quark distributions are clearly flavor asymmetric.
Inherent in this assumption is that the content of the sea is 
independent of the valence quark's composition. Therefore,
the proton and neutron were expected to have identical sea-quark distributions.
The assumption of flavor symmetry was not based on any known physics, and 
it remained to be tested by experiments. Neutrino-induced charm production
experiments~\cite{abrom82,conrad98}, which are 
sensitive to the $s \to c$ process, provided strong
evidences that the strange-quark content of the nucleon is only about half
of the up or down sea quarks. This flavor asymmetry was attributed to the
much heavier mass for strange quark compared to the up and down quarks. 
The mass for the up and down quarks being very similar suggests that 
the nucleon sea should be nearly
up-down symmetric. A direct method
to check this assumption is to compare the sea in the neutron 
to that in the proton by measuring the Gottfried integral 
in DIS, as discussed next.

\subsection{\it Gottfried Sum Rule}
In 1967, Gottfried studied electron-proton scattering with the assumption
that the proton consists of three constituent quarks~\cite{gott67}. He 
showed that the total
electron-proton cross section (elastic plus inelastic) is identical to 
the Rutherford scattering from a point charge. Gottfried derived a sum rule
\begin{eqnarray}
I^p_2 = \int_0^1 F^p_2 (x,Q^2)/x~ dx = \sum_i (Q^{p}_{i})^2 = 1,
\label{eq:2.2.1}
\end{eqnarray}
where $Q^{p}_{i}$ is the charge of the $i$th quark in the proton.
Gottfried expressed great skepticism that this sum rule would be confirmed by
the forthcoming SLAC experiment by stating ``{\it I think Prof. Bjorken and
I constructed the sum rules in the hope of destroying 
the quark model}''~\cite{bj67}.
Indeed, Eq.~\ref{eq:2.2.1} was not confirmed by the 
experiments, not because of the
failure of the quark model, but because of the presence of quark-antiquark sea.
In fact, the total number of the sea partons being infinite makes $I^p_2$ 
diverge. A closely related sum rule, now called the Gottfried Sum Rule (GSR),
avoids this problem by considering the difference
of the proton and neutron cross sections, namely,
\begin{eqnarray}
I^p_2 - I^n_2 = \int_0^1 [F^p_2 (x,Q^2)-F^n_2 (x,Q^2)]/x~ dx 
= \sum_i [(Q^{p}_{i})^2 - (Q^{n}_{i})^2] = 1/3.
\label{eq:2.2.2}
\end{eqnarray}
In deriving Eq.~\ref{eq:2.2.2}, it was assumed 
that the sea quarks in the proton and 
neutron are identical. Regarding the Gottfried Sum Rule,
Kuti and Weisskopf stated in their paper~\cite{kuti71} ``{\it 
This very simple and 
definite sum rule is based upon our most radical simplification: the 
assumption that the core carries vacuum quantum numbers so that the 
isotopic spin of the nucleon is completely carried by the valence quarks}".

Soon after the discovery of scaling in electron-proton DIS,
electron-deuterium scattering experiments were carried out to
extract the electron-neutron cross sections. The comparison of
$e-p$ with $e-n$ data was very important for distinguishing early competing
theoretical models~\cite{friedman}. These data also allowed a
first evaluation~\cite{bloom70} of the Gottfried integral
in 1970. The
first result for the Gottfried integral was 0.19, considerably less than 1/3.
As the data only covered $x > 0.08$, it was assumed that 
$F^p_2 - F^n_2$ follows Regge behavior (proportional to $x^{1/2}$)
in the unmeasured small-$x$ region. Due to the $1/x$ factor in the integrand, 
the small-$x$ region could have potentially large contributions to
the Gottfried integral. Moreover, it was not clear if $F^p_2 - F^n_2$ would
indeed follow the Regge behavior at small $x$, and if so, at what value of
$x$ would it set in. By 1973, new data were available down to $x = 0.05$ 
and the Gottfried
integral was evaluated to be 0.28~\cite{bloom73}, considerably larger
than the first result. 
It should be pointed out that these data were taken at relatively low
values of $Q^2$. Furthermore, $Q^2$ varied as a function
of $x$. 

Although the large systematic errors associated with 
the unmeasured small-$x$ region prevented
a sensitive test of the GSR, Field and 
Feynman~\cite{field77} nevertheless
interpreted the early SLAC data as a strong indication that GSR is
violated and that the $\bar u$ and $\bar d$ distributions in the proton 
are different.
The relationship between the Gottfried integral and the 
$\bar d/ \bar u$ asymmetry is clearly seen in the parton model, namely,
\begin{eqnarray}
\int_0^1 [F^p_2 (x,Q^2)-F^n_2 (x,Q^2)] /x~ dx 
= {1 \over 3} + {2 \over 3}
\int_0^1 [\bar u(x,Q^2) - \bar d(x,Q^2)] dx.
\label{eq:2.2.2.1}
\end{eqnarray}
Equation~\ref{eq:2.2.2.1} clearly shows that the early SLAC data on GSR
implied $\bar d > \bar u$, at least for certain region of $x$.
Field and Feynman further suggested that Pauli blocking
from the valence quarks would inhibit the $\bar u u$ sea more than the
$\bar d d$ sea, hence creating an asymmetric nucleon sea.

\begin{figure}[tb]
\begin{center}
\begin{minipage}[t]{11 cm}
\epsfig{file=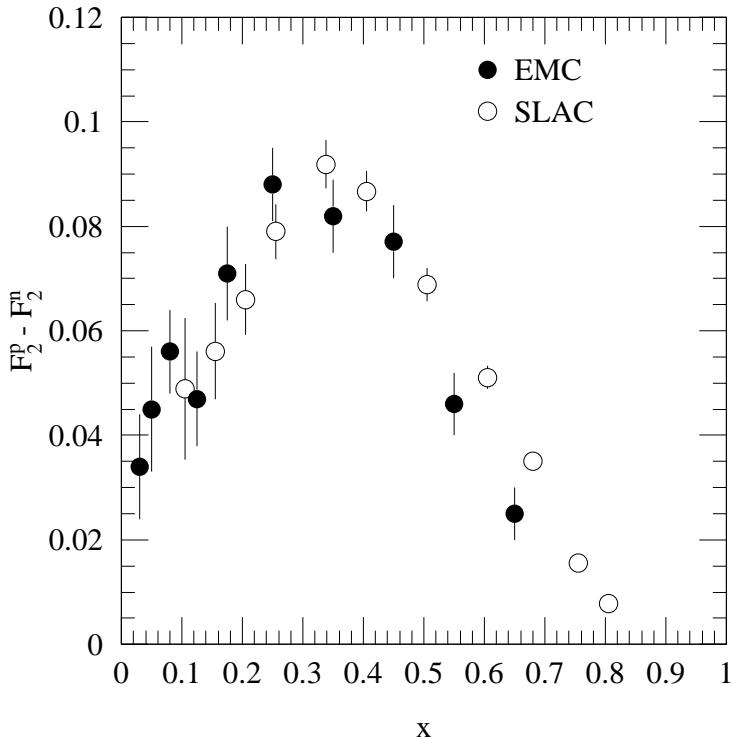,scale=1.2}
\end{minipage}
\begin{minipage}[t]{16.5 cm}
\caption{EMC~\cite{emc87} and SLAC~\cite{bodek79} measurements                  
of $F^p_2 - F^n_2$.}                                                            
\label{fig:2.2.1}
\end{minipage}
\end{center}
\end{figure}

The SLAC DIS experiments were followed by several muon-induced DIS experiments
at Fermilab and at CERN. Using high energy muon beams, these
experiments reached much larger values of $Q^2$ and they
clearly observed~\cite{wate75,chang75,chio79} the scaling-violation 
phenomenon in DIS. 
The Gottfried integral was also evaluated in muon DIS 
experiments~\cite{emc87,bcdms90}. Figure~\ref{fig:2.2.1}
compares the data from the European Muon 
Collaboration (EMC)~\cite{emc87} with earlier
electron data from SLAC~\cite{bodek79}. The coverages 
in $x$ are similar in these two
experiments, even though
the $Q^2$ values covered by the EMC are much larger.
Figure~\ref{fig:2.2.1} shows that $F^p_2 -F^n_2$ from EMC tend to shift towards 
smaller $x$ relative to the SLAC data, in qualitative agreement 
with the QCD $Q^2$-evolution. 
The Gottfried integral 
determined from the EMC experiment is $0.235 + 0.110 - 0.099$, consistent
with the result from SLAC, but still lower than 1/3. 

Despite the fact that all measurements of the 
Gottfried integral consistently showed
a value lower than 1/3, the large systematic errors prevented a
definitive conclusion. As a result, all 
parametrizations~\cite{do84,ehlq84,dflm88,mrs88,abfow89}
of the parton distributions based on global fits to existing data before 1990 
assumed a symmetric $\bar u$, $\bar d$
sea. As discussed next, more compelling evidences for an asymmetric 
light-quark sea were provided by results from new experiments.
 
\subsection{\it Drell-Yan Process}
The first high-mass dilepton production experiment~\cite{chris70} 
was carried out at the 
AGS in 1969, soon after scaling was discovered at SLAC.
Drell and Yan~\cite{dy71} interpreted the 
data within the parton model, in which 
a quark-antiquark pair annihilate into a virtual
photon subsequently decaying into a lepton pair. This simple model was
capable of explaining several pertinent features of the data, including the
overall magnitude of the cross sections, the scaling behavior of the cross
sections, and the polarization of the virtual photons. The high-mass continuum
lepton-pair production is therefore called the Drell-Yan (DY) process.

Since the underlying mechanism for the DY process involves the annihilation
of a quark with an antiquark, it is not surprising that this process can
be used to probe the antiquark contents of the beam or target hadrons.
In the parton model, the DY cross section is given by
\begin{eqnarray}
{d^2\sigma\over dM^2dx_F}={4\pi\alpha^2\over 9M^2s}{1\over (x_1+x_2)}\sum_a
e_a^2[q_a(x_1)
\bar q_a(x_2)+\bar q_a(x_1)q_a(x_2)]. 
\label{eq:2.3.1}
\end{eqnarray}
Here $q_a(x)$ and $\bar q_a(x)$ are the quark and 
antiquark parton distributions of 
the two colliding hadrons evaluated at the momentum fraction $x$.
The sum is over quark flavors.  
In addition, one has the kinematic relations,
\begin{eqnarray}
& &\tau\equiv x_1x_2 = M^2/s,\nonumber \\
& &x_F = x_1-x_2, 
\label{eq:2.3.2}
\end{eqnarray}
where $M$ is the invariant mass of the lepton pair and $s$ is the square 
of the center-of-mass energy. The cross section is 
proportional to $\alpha^2$, indicating its electromagnetic character. 

Equation~\ref{eq:2.3.1} shows that the 
antiquark distribution enters as a multiplicative
term in the DY cross section rather than an additive term in the
DIS cross section. Hence, the antiquark
distributions can be sensitively determined in the DY experiments. The 
dimuon data from the FNAL E288, in which the $\Upsilon$ resonances were 
discovered~\cite{herb77}, were analysed~\cite{kaplan77} to 
extract the sea quark distribution
in the nucleon. By assuming a flavor-symmetric nucleon sea, namely,
$\bar u(x) = \bar d(x) = \bar s(x) = Sea(x)$, the dimuon
mass distribution obtained in 400 GeV proton-nucleus interaction was  
described by $xSea(x) = 0.6 (1-x)^{10}$.
In a later analysis~\cite{ito81} taking into account
additional data at 200 and 300 GeV,
the E288 collaboration found that a much better fit could be obtained with an
asymmetric sea, namely,
\begin{eqnarray}
&\bar u(x) = (1-x)^{3.48} \bar d(x),~ ~ ~ ~\bar s(x) = (\bar u(x) + \bar d(x))/4.
\label{eq:2.3.3}
\end{eqnarray}
The need for an asymmetric $\bar u$ and $\bar d$ was also revealed 
in the E288 $d\sigma/dy$ DY data~\cite{ito81} at $y = 0$, where $y$ is 
the center-of-mass rapidity. For $p + A$ collision, the slope of 
$d\sigma/dy$ at $y = 0$ is expected to
be positive due to the excess of $u$ over $d$ valence quarks in the proton.
The E288 data showed that the slopes are indeed positive, 
but larger than expected from a flavor symmetric sea.
A surplus of $\bar d$ 
over $\bar u$ in the proton sea
would lead to more positive slope in agreement with the data~\cite{ito81}.

The FNAL E439 collaboration~\cite{smith81} studied high 
mass dimuons produced in $p + W$ interaction at 400 GeV.
Their spectrometer covered a considerably larger range
in $x_F$ than E288. They again found that an asymmetric sea, 
$\bar u(x) = (1-x)^{2.5} \bar d(x) $, could well describe their data.

With all the tantalizing evidence for an asymmetric sea
from DIS and DY experiments, it is curious
that all global analyses~\cite{do84,ehlq84,dflm88,mrs88,abfow89} 
of parton distributions in the 1980's still
assumed a symmetric light-quark sea. This probably reflected the 
reluctance to adopt an unconventional description of the nucleon sea without
compelling and unambiguous experimental evidence. As discussed in the
next Section, such evidence became available in the 1990's.

\section{Recent Experimental Developments}
\subsection{\it NMC Measurements of the Gottfried Integral}
After the discovery of the so-called `EMC effect'~\cite{emc83} 
which showed that the parton
distributions in heavy nuclei are different from that in the deuteron, the EMC
detector system was modified by the New Muon Collaboration (NMC) at CERN to
study in detail the EMC effect. Special emphases were placed on the
capability to reach the 
low-$x$ region where the `shadowing effect'~\cite{emc88} 
is important, and to measure cross section ratios accurately~\cite{nmc90}.
By implementing a `small angle' trigger which extended the 
scattering angle coverage down to 5 mrad, the lowest value of $x$ reached 
by NMC was $\sim 0.001$. The NMC also placed two targets in the
muon beam, allowing DIS data from two different targets to be recorded
simultaneously, thus greatly reducing the beam flux normalization uncertainty.
To account for the different geometric acceptances for events originating from
the two targets, half of the data were taken using a target 
configuration where the locations of the two targets were interchanged.

\begin{figure}[tb]
\begin{center}
\begin{minipage}[t]{9 cm}
\epsfig{file=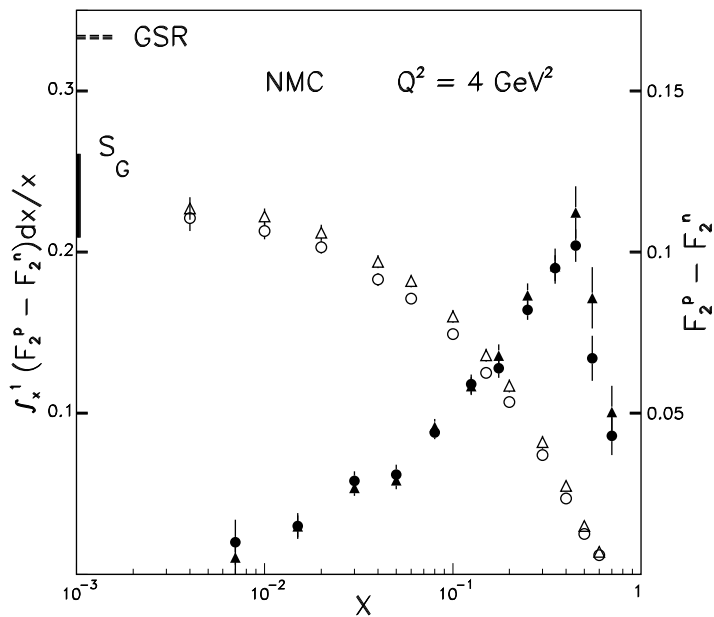,scale=0.5}
\end{minipage}
\begin{minipage}[t]{16.5 cm}
\caption{NMC measurements of $F^p_2 - F^n_2$ (solid data points)
and the Gottfried integral (open data points). The triangular data
points correspond to results published in 1991~\cite{nmc91}, while
the circular data points represent a more recent analysis in
1994~\cite{nmc94}. The extrapolated value of Gottfried integral ($S_G$)
and the expected GSR value are also indicated.}
\label{fig:3.1.1}
\end{minipage}
\end{center}
\end{figure}

The upgraded NMC detectors allowed a definitive study~\cite{nmc95} 
of the shadowing 
effect at low $x$. Moreover, they enabled a much more accurate 
determination of the Gottfried integral. Figure~\ref{fig:3.1.1} 
shows the $F^p_2 - F^n_2$ 
reported by the NMC in 1991~\cite{nmc91}, in which 
the smallest $x$ reached (0.004) was
significantly lower than in previous experiments. 
Taking advantage of their accurate measurements of the $F^n_2/F^p_2$ 
ratios, NMC used the following expression to evaluate $F^p_2 - F^n_2$, namely,
\begin{equation}
F^p_2 - F^n_2 = F^d_2~(1 - F^n_2/F^p_2)/(1 + F^n_2/F^p_2).
\label{eq:3.1.1}
\end{equation}
The ratio $F^n_2/F^p_2 \equiv F^d_2/F^p_2 - 1$ was determined from NMC's
$F^d_2/F^p_2$ measurement, while $F^d_2$ ($F^d_2 = F^p_2 + F^n_2$) was
taken from a fit to previous DIS experiments 
at SLAC~\cite{slac92}, Fermilab~\cite{chio79}, and
CERN~\cite{bcdms90a,emc90}. The value of the Gottfried integral 
for the measured region at $Q^2$ = 4 GeV$^2$ is 
$S_G(0.004 - 0.8) = 0.227 \pm 0.007(stat) \pm 0.014(syst)$.
The contribution to $S_G$ from $x > 0.8$ was estimated to be
$0.002 \pm 0.001$. Assuming that $F^p_2 - F^n_2$ at $x < 0.004$ behaves
as $ax^b$, NMC estimated $S_G(0 - 0.004) = 0.011 \pm 0.003$. Summing the 
contributions from all regions of $x$, NMC obtained 
$S_G = 0.240 \pm 0.016$,
which was significantly below 1/3. This represented the best evidence thus far
that the GSR was indeed violated.

In 1994, NMC reevaluated~\cite{nmc94} 
the Gottfried integral using a new $F^d_2$
parametrization from Ref.~\cite{nmc92} and newly determined 
values of $F^d_2/F^p_2$.
The new $F^d_2$ parametrization include data from 
SLAC~\cite{slac92}, BCDMS~\cite{bcdms90a}, as well as
NMC's own measurement~\cite{nmc92}. The new NMC 
results for $F^p_2 - F^n_2$ are shown 
in Fig.~\ref{fig:3.1.1}. Note that the 1994 
values are slightly larger (smaller) than
the 1991 values at small (large) $x$. 
The new evaluation gave $S_G(0.004 - 0.8) = 0.221 \pm 0.008(stat) 
\pm 0.019(syst)$, and the Gottfried integral became
\begin{equation}
S_G = 0.235 \pm 0.026.
\label{eq:3.1.2}
\end{equation}
This  value is consistent with the earlier number. 
The new systematic error is larger than the old one, reducing somewhat 
the overall significance of the NMC measurement. Nevertheless, the violation 
of the GSR is established at a $4 \sigma$ level.

More recently, NMC published their final 
analysis of $F^d_2/F^p_2$~\cite{nmc97}
and $F^d_2$~\cite{nmc95a,nmc97a}. This analysis 
included the 90 and 280 GeV data 
taken in 1986 and 1987, as well as the 1989 data at 120, 200 and 280 GeV.
The 1989 data were not used in the 
earlier evaluations~\cite{nmc91,nmc94} of the Gottfried
integral. Based on these new data, NMC reported 
$S_G(0.004-0.8) = 0.2281 \pm 0.0065(stat)$ 
at $Q^2$ = 4 GeV$^2$~\cite{nmc97}. 
This agrees within statistical errors with previous 
NMC results~\cite{nmc91,nmc94}.

QCD corrections to various parton-model sum rules have been reviewed
recently~\cite{hinch96}. The $\alpha_s$ and
$\alpha_s^2$ corrections to the Gottfried integral have been 
calculated~\cite{ross79,kotaev96} and found to be very small (roughly
0.4 \% each at $Q^2$ = 4 GeV$^2$). Therefore, QCD corrections can not
account for the large violation of GSR found by the NMC. Although perturbative
QCD predicts a weak $Q^2$ dependence for the Gottfried integral, it has been 
suggested~\cite{forte94a, forte94b} that due to the non-perturbative origin of
the $\bar d$, $\bar u$ asymmetry the $Q^2$ dependence of 
the Gottfried integral will be anomalous between 1 and 5 GeV$^2$.
This interesting suggestion remains to be tested by DIS
experiments.

\subsection{\it E772 Drell-Yan Experiment}
The main goal of the Fermilab experiment E772 was to 
examine the origin of the EMC effect. Among the many theoretical models 
which explain the EMC effect~\cite{gee95}, the 
pion-excess model~\cite{pionex83} predicted a large
enhancement of antiquark content due to the presence of additional meson
cloud in heavy nuclei. This prediction could be readily tested by measuring the
nuclear dependence of proton-induced DY cross sections. Using an 800 GeV proton 
beam, the DY cross section ratios of a number of nuclear targets 
($C, Ca, Fe, W$)
were measured~\cite{e772a} relative to deuterium 
with high statistics over the region
$0.04 < x_2 < 0.35$. The enhancement of antiquark contents predicted by 
the pion-excess model was not observed, and the E772 results
were in good agreement with the prediction of the 
rescaling model~\cite{close85}.

Information on the $\bar d/\bar u$ asymmetry has also been extracted from the
E772 data~\cite{e772b}. At $x_F > 0.1$, the dominant 
contribution to the proton-induced
DY cross section comes from the annihilation of $u$ quark in the projectile
with the $\bar u$ quark in the target nucleus. It follows that the DY cross 
section (per nucleon) ratio of a non-isoscalar target 
(like $^1H, Fe, W$) over an isoscalar
target (like $^2H, ^{12}C, ^{40}Ca$) is given as
\begin{eqnarray}
R_A(x_2) \equiv \sigma_A(x_2)/\sigma_{IS}(x_2) \approx
1 + [(N-Z)/A] [(1 - \bar u(x_2)/\bar d (x_2))/(1 + \bar u(x_2)/\bar d(x_2))],
\label{eq:3.2.1}
\end{eqnarray}
where $x_2$ is the Bjorken-$x$ of the target partons, IS stands for isoscalar,
and $N, Z$ and $A$ refer to the non-isoscalar target. The $(N-Z)/A$ factor 
in Eq.~\ref{eq:3.2.1} shows that the largest sensitivity to the $\bar d/ \bar u$
could be obtained with a measurement of $\sigma(p+p)/\sigma(p+d)$.
Nevertheless, for $W$ target $(N-Z)/A = 0.183$ and the E772 
$\sigma_W/\sigma_{IS}$ data could be used to study the $\bar d, \bar u$
asymmetry.

\begin{figure}[tb]
\begin{center}
\begin{minipage}[t]{11 cm}
\epsfig{file=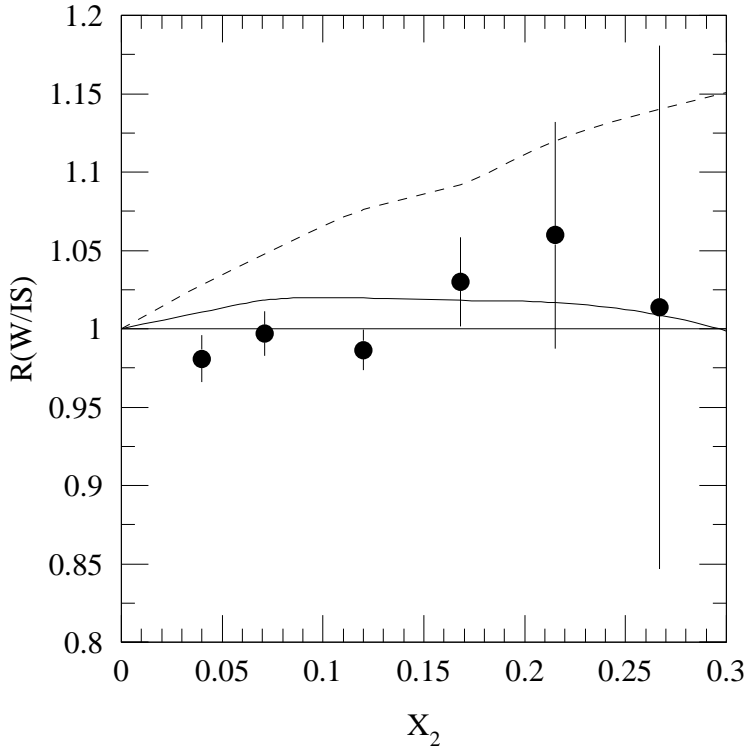,scale=1.2}
\end{minipage}
\begin{minipage}[t]{16.5 cm}
\caption{The ratios of $W$ over isoscalar targets DY cross sections
from E772~\cite{e772b}. The dashed curve is a calculation using the
$\bar d / \bar u$ asymmetric parton distributions suggested
in Ref.~\cite{es91}. The solid curve corresponds to a calculation
using the MRST distribution functions.}
\label{fig:3.2.1}
\end{minipage}
\end{center}
\end{figure}

Figure~\ref{fig:3.2.1} shows the E772 DY cross section ratios from 
the neutron-rich $W$ target over the isoscalar targets, $^2H$ and $^{12}C$.
Corrections have been applied to the two data points at $x_2 < 0.1$ to 
account for the nuclear shadowing effects in $C$ and $W$~\cite{e772b}. 
The E772 data 
provided some useful constraints on the $\bar d/\bar u$ asymmetry. 
In particular, some early parametrizations~\cite{es91,ehq92} 
of large $\bar d/\bar u$ 
asymmetry were ruled out. Despite the relatively large 
error bars, Figure~\ref{fig:3.2.1} shows $R > 1.0$ at $x_2 > 0.15$, 
which is consistent with $\bar d > \bar u$ in this region. 

\begin{figure}[tb]
\begin{center}
\begin{minipage}[t]{11 cm}
\epsfig{file=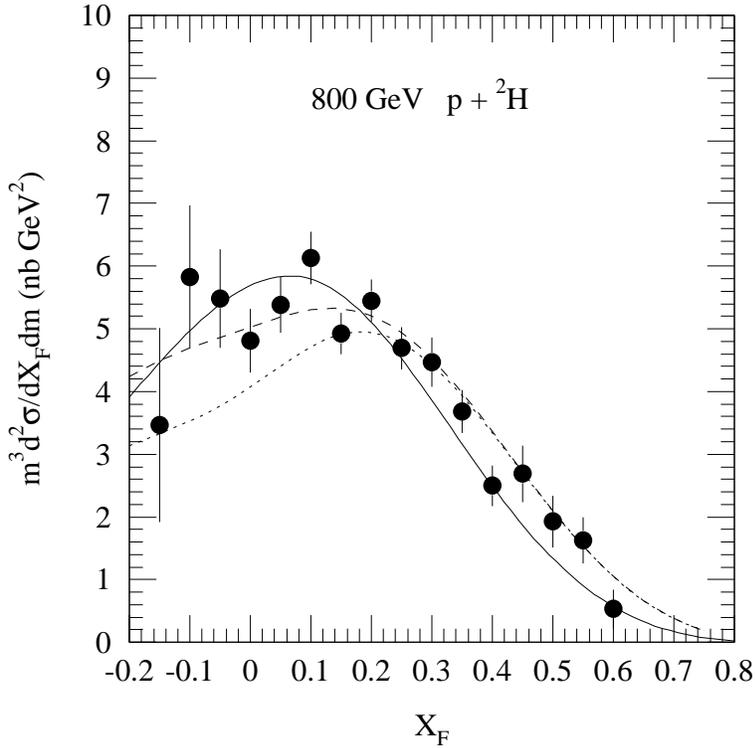,scale=1.2}
\end{minipage}
\begin{minipage}[t]{16.5 cm}
\caption{The $p + ^2H$ DY differential cross sections from E772~\cite{e772b}.
The dotted (dashed) curve is a calculation using the parton
distributions from Ref.~\cite{es91} with (without)
$\bar d / \bar u$ asymmetry. The solid curve uses the MRST
distribution functions. These are leading-order calculations normalized to
the data.}
\label{fig:3.2.2}
\end{minipage}
\end{center}
\end{figure}

E772 collaboration also presented the DY differential cross sections for
$p+d$ at $\langle M_{\mu\mu}\rangle = 8.15$ GeV. As shown in 
Fig.~\ref{fig:3.2.2},
the DY cross sections near $x_F = 0$ are sensitive to $\bar d/\bar u$
and the data disfavor a large $\bar d/\bar u$ asymmetry. Figure~\ref{fig:3.2.2}
also shows that a recent parton distribution function, 
MRST~\cite{mrst}, which has modest 
$\bar d/\bar u$ asymmetry, is capable of describing the $p+d$ differential 
cross sections well (see Section 3.6).

While the E772 data provide some useful constraints on the values 
of $\bar d/\bar u$, it is clear that a measurement of 
$\sigma_{DY}(p+d)/\sigma_{DY}(p+p)$ is highly desirable. The $(N-Z)/A$ 
factor is now equal to $-1$, indicating a large improvement in the 
sensitivity to $\bar d/\bar u$. Moreover, the uncertainty arising from
nuclear effects would be much reduced. It has also been pointed 
out~\cite{kumano95} that $\bar d/\bar u$ asymmetry in the 
nucleon could be significantly modified
in heavy nuclei through recombination of partons from different nucleons.
Clearly, an ideal approach is to first determine $\bar d/\bar u$ in the
nucleon before extracting the $\bar d/\bar u$ information in heavy nuclei.

\subsection{\it NA51 Drell-Yan Experiment}
Following the suggestion of Ellis and Stirling~\cite{es91},
the NA51 collaboration at CERN carried out the first dedicated dimuon
production experiment to study the flavor structure of the 
nucleon sea~\cite{na51}.
Using a 450 GeV proton beam, roughly 2800 and 3000 dimuon events with 
$M_{\mu\mu} > 4.3$ GeV have been recorded, respectively, for $p+p$ and
$p+d$ interaction. The spectrometer setting covers the kinematic region
near $y = 0$. At $y = 0$, the asymmetry parameter, $A_{DY}$, is given as
\begin{eqnarray}
A_{DY} \equiv {\sigma^{pp} -\sigma^{pn}\over \sigma^{pp} + \sigma^{pn}}
\approx {(4\lambda_V - 1)(\lambda_s - 1) + (\lambda_V - 1)(4\lambda_s - 1)\over
(4\lambda_V + 1)(\lambda_s + 1) + (\lambda_V + 1)(4\lambda_s + 1)},
\label{eq:3.2.2}
\end{eqnarray}
where $\lambda_V = u_V/d_V$ and $\lambda_s = \bar u/\bar d$. 
In deriving Eq.~\ref{eq:3.2.2}, the negligible sea-sea annihilation 
was ignored and the validity of charge symmetry was assumed. At $x = 0.18$,
$\lambda_V \approx 2$ and according to Eq.~\ref{eq:3.2.2} $A_{DY} = 0.09$ for
a symmetric sea, $\lambda_s = 1$.
For an asymmetric $\bar d > \bar u$ sea, $A_{DY}$ would be less than 0.09.

From the DY cross section ratio, $\sigma^{pp}/\sigma^{pd}$, NA51
obtained $A_{DY} = -0.09 \pm 0.02 (stat) \pm 0.025 (syst)$. This then led
to a determination of $\bar u/\bar d = 0.51 \pm 0.04 (stat) \pm (syst)$
at $x = 0.18$ and $\langle M_{\mu\mu}\rangle = 5.22$ GeV. This 
important result established
the asymmetry of the quark sea at a single value of $x$. What remained
to be done was to map out the $x$-dependence of this asymmetry. This was
subsequently carried out by the Fermilab E866/NuSea 
collaboration, as discused next.

\subsection{\it E866 Drell-Yan Experiment}
At Fermilab, a DY experiment (E866/NuSea) aimed at a higher statistical
accuracy with a much wider kinematic coverage than the NA51 experiment was
recently completed~\cite{e866}. This experiment 
measured the DY muon pairs from 800
GeV proton interacting with liquid deuterium and hydrogen targets.
A proton beam with up to $2 \times 10^{12}$ 
protons per 20 s spill bombarded one 
of three identical 50.8-cm long cylindrical target flasks containing either
liquid hydrogen, liquid deuterium or vacuum. The targets alternated every 
few beam spills in order to minimize time-dependent systematic effects.
The dimuons accepted by a 3-dipole magnet spectrometer were detected by
four tracking stations. An integrated flux of $1.3 \times 10^{17}$ protons
was delivered for this measurement. 

\begin{figure}[tb]
\begin{center}
\begin{minipage}[t]{11 cm}
\epsfig{file=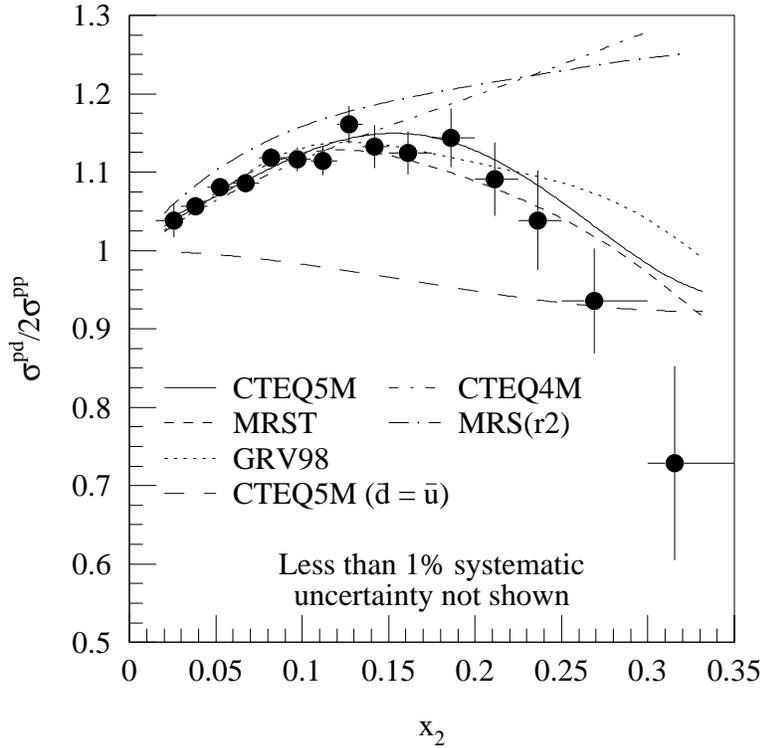,scale=1.2}
\end{minipage}
\begin{minipage}[t]{16.5 cm}
\caption{The ratio $\sigma^{pd}/2\sigma^{pp}$ of Drell-Yan cross
sections versus $x_{2}$ for E866. The curves are next-to-leading order
calculations, weighted by acceptance, of the Drell-Yan cross section
ratio using various parton distributions. The CTEQ4M and MRS(R2) 
parton distributions existed prior to the E866 data, while the other 
distributions came after the E866 results were obtained.
In the lower
CTEQ5M curve $\bar{d} - \bar{u}$ has been arbitrarily set to 0 as
described in the text. The errors are statistical only.}
\label{fig:3.4.1}
\end{minipage}
\end{center}
\end{figure}

Over 330,000 DY events were recorded in E866, using three different 
spectrometer settings which covered the regions of low, intermediate and high 
mass muon pairs. The data presented here are from the analysis of 
the full data set~\cite{towell}. 
These data are in good qualitative agreement with the high-mass
data published earlier~\cite{e866}. The DY cross section ratio
per nucleon for $p + d$ to that for $p + p$ is 
shown in Fig.~\ref{fig:3.4.1} as a function of 
$x_2$. The acceptance of the spectrometer
was largest for $x_F = x_1 - x_2 > 0$. In this kinematic regime the DY
cross section is dominated by the annihilation of a beam quark with a target
antiquark. To a very good approximation the DY cross section ratio
at positive $x_F$ is given as 
\begin{equation}
\sigma_{DY}(p+d)/2\sigma_{DY}(p+p) \simeq
(1+\bar d(x_2)/\bar u(x_2))/2.
\label{eq:3.3}
\end{equation}
In the case that $\bar d = \bar u$, the ratio is 1.
Figure~\ref{fig:3.4.1} shows that the DY cross section per nucleon for 
$p + d$ clearly exceeds $p + p$, and it indicates an excess of $\bar d$ 
with respect to $\bar u$ over an appreciable range in $x_2$.

Figure~\ref{fig:3.4.1} also compares the data with next-to-leading order (NLO)
calculations of the cross section ratio using the CTEQ4M~\cite{cteq} and
the MRS(R2)~\cite{mrs} parton distributions. The data are in reasonable 
agreement with the predictions for $x_{2}<0.15$. Above
$x_{2}=0.15$ the data lie well below the CTEQ4M and the MRS(R2) values.
Following the publication of the E866 high-mass data~\cite{e866}, 
several new parametrizations for parton distributions have been
put forward. In Fig.~\ref{fig:3.4.1} we compare the full E866 data with the
calculations using CTEQ5M~\cite{cteq5}, MRST~\cite{mrst}, and
GRV98~\cite{grv98}. Also shown in Fig.~\ref{fig:3.4.1} is a calculation 
using a modified CTEQ5M parton distributions. 
The modified CTEQ5M parton distributions, in which the
$\bar{d}+\bar{u}$ parametrization was maintained but $\bar d$ was set 
identical to $\bar u$, were used to illustrate the cross section ratio
expected for a symmetric $\bar d/ \bar u$ sea. The E866 data clearly show 
that $\bar d \ne \bar u$. 

\begin{figure}[tb]
\begin{center}
\begin{minipage}[t]{11 cm}
\epsfig{file=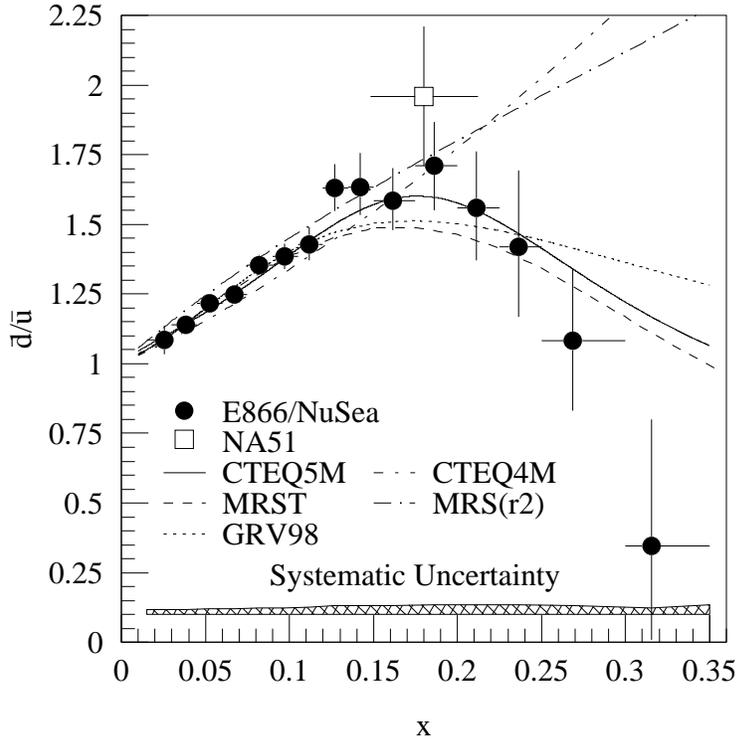,scale=1.2}
\end{minipage}
\begin{minipage}[t]{16.5 cm}
\caption{The ratio of $\bar{d}/\bar{u}$ in the proton as a function
of $x$ extracted from the Fermilab E866 cross section ratio. The
curves are parametrizations of various parton distribution functions.
The error bars indicate statistical errors only.  
Also shown is the result
from NA51, plotted as an open box.}
\label{fig:3.4.2}
\end{minipage}
\end{center}
\end{figure}

Using an iterative procedure described in \cite{e866,towell,peng98}, 
values for $\bar d/ \bar u$ were extracted by the E866 collaboration at
$Q^2 = 54$ GeV$^2$ and shown in Fig.~\ref{fig:3.4.2}.
At $x < 0.15$, $\bar d/\bar u$ increases linearly with $x$ and is in
good agreement with the CTEQ4M and MRS(R2) parametrization. 
However, a distinct feature of the data, not seen in either 
parametrization of the parton distributions, is the
rapid decrease towards unity of the $\bar{d}/\bar{u}$ ratio beyond
$x_{2}=0.2$\@. Figure~\ref{fig:3.4.2} shows that the most recent
parton distribution parametrizations (MRST, CTQ5M, GRV98) adequately
describe the E866 data. 
The result from NA51 is also shown in Figure~\ref{fig:3.4.2}
for comparison.

\begin{figure}[tb]
\begin{center}
\begin{minipage}[t]{11 cm}
\epsfig{file=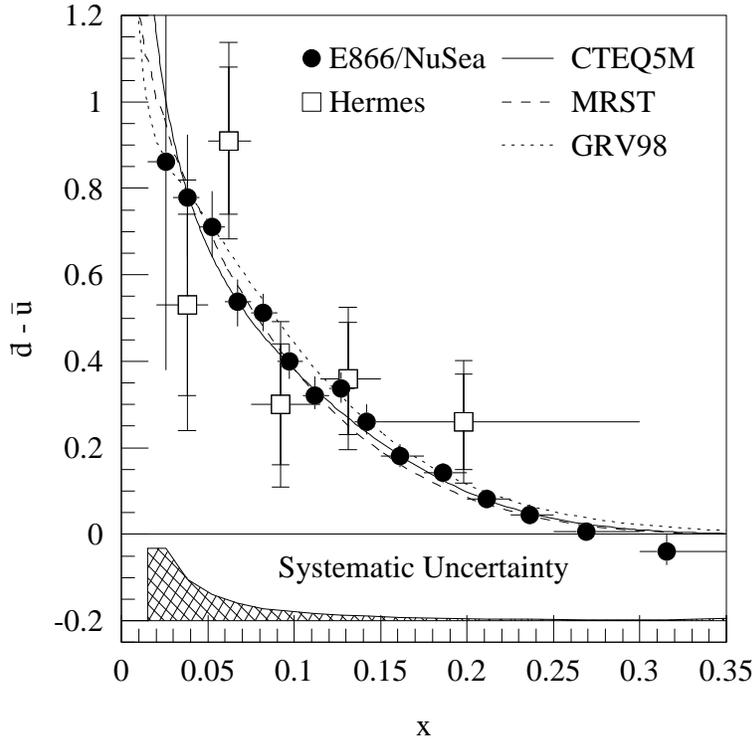,scale=1.2}
\end{minipage}
\begin{minipage}[t]{16.5 cm}
\caption{Comparison of the E866~\cite{e866} $\bar d - \bar u$ results at $Q^2$ =
54 GeV$^2$ with the parametrizations of various parton distribution
functions. 
The data from HERMES~\cite{hermes98} are also shown.}
\label{fig:3.4.3}
\end{minipage}
\end{center}
\end{figure}

The $\bar d / \bar u$ ratios measured in E866, together with the
CTEQ5M values for $\bar d + \bar u$, were used to obtain
$\bar d - \bar u$ over the region $0.02 < x < 0.345$
(Fig.~\ref{fig:3.4.3}). As a flavor non-singlet quantity,
$\bar d(x) - \bar u(x)$ is decoupled from the effects of the gluons and
is a direct measure of the contribution from non-perturbative processes, 
since perturbative processes cannot cause a significant $\bar d$, $\bar u$
difference. From the results shown in Fig.~\ref{fig:3.4.3}, one can
obtain an independent determination~\cite{towell} of the integral
of $\bar d(x) - \bar u(x)$ and compare it with the NMC result 
(Eq.~\ref{eq:3.1.2}). E866 obtains a value 
$\int_0^1 \left[\bar d(x) - \bar u(x)\right] dx = 0.118 \pm 0.012$,
which is $4/5$ the value deduced by NMC~\cite{nmc94}.

\begin{figure}[tb]
\begin{center}
\begin{minipage}[t]{11 cm}
\epsfig{file=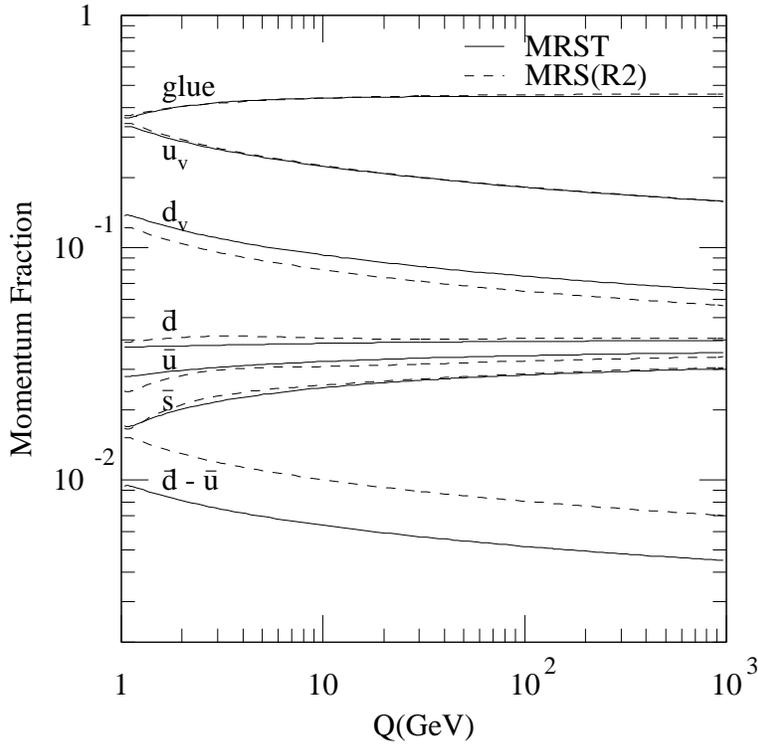,scale=1.2}
\end{minipage}
\begin{minipage}[t]{16.5 cm}
\caption{$Q$-dependence of the proton's momentum carried by
various partons calculated using the MRS(R2) and MRST parton
distribution functions.}
\label{fig:3.4.4}
\end{minipage}
\end{center}
\end{figure}

The E866 data also allow the first determination~\cite{peng98} 
of the difference of
the momentum fraction
carried by $\bar d$ and $\bar u$. 
One obtains 
$\int_{0.02}^{0.345} x
\left[\bar d(x) - \bar u(x)\right] dx = 0.0065 \pm 0.0010$ at 
$Q^2$ = 54 GeV$^2$.
If CTEQ4M is used to estimate
the contributions from the unmeasured $x$ regions, one finds that 
$\int_0^1 x \left[\bar d(x) - \bar u(x)\right] dx = 0.0075 \pm 0.0011$.
Unlike the integral of $\bar
d(x) - \bar u(x)$, the momentum integral is $Q^2$ dependent and
decreases as $Q^2$ increases. The $Q^2$ dependence of the momentum
fraction carried by various partons are shown in Fig.~\ref{fig:3.4.4}.
The calculation uses both the MRS(R2) and the new MRST~\cite{mrst} 
parton distributions for comparison. Figure~\ref{fig:3.4.4}
shows that the momentum fractions carried by the sea quarks increase
with $Q^2$. In contrast, the difference of the momentum fraction carried
by up and down sea quarks decreases with $Q^2$.

\subsection{\it HERMES Semi-Inclusive Experiment}
It has been recognized for some time that semi-inclusive DIS could be
used to extract the flavor dependence of the valence
quark distributions~\cite{grz73}. From quark-parton model, the 
semi-inclusive cross
section, $\sigma^h_N$, for producing a hadron on a nucleon is given by
\begin{eqnarray}
{1 \over \sigma_N(x)} {d\sigma^h_N(x,z)\over dz} = 
{\sum_i e^2_i f_i(x) D^h_i(z)\over\sum_i e^2_i f_i(x)},
\label{eq:3.5.1}
\end{eqnarray}
where $D^h_i(z)$ is the fragmentation function signifying the probability for
a quark of flavor $i$ fragmenting into a hadron $h$ carrying
a fraction $z$ of the initial quark momentum. $e_i$ and $f_i$ are the charge
and the distribution function of quark $i$, and $\sigma_N(x)$ is the
inclusive DIS cross section. Assuming
charge symmetry for the fragmentation functions and parton distribution
functions, one can derive the relationship
\begin{eqnarray}
{d_v(x)\over u_v(x)} ={4 R^\pi(x) + 1 \over 4 + R^\pi (x)}, 
\label{eq:3.5.2}
\end{eqnarray}
where
\begin{eqnarray}
R^\pi(x) = (d\sigma_n^{\pi^+}(x,z)/dz -d\sigma_n^{\pi^-}(x,z)/dz)/
(d\sigma_p^{\pi^+}(x,z)/dz -d\sigma_p^{\pi^-}(x,z)/dz).
\label{eq:3.5.3}
\end{eqnarray}
Based on a large number of semi-inclusive charged-hadron events in
muon DIS from hydrogen and deuterium targets, EMC extracted~\cite{ashman91}
the values
of $d_v(x)/u_v(x)$ over the range $0.028 < x < 0.66$. The EMC result agrees
with neutrino measurements~\cite{bebc84,cdhs84}, and 
it demonstrates the usefulness of semi-inclusive measurements
for extracting valence quark distributions.

Soon after the report of GSR violation by the NMC, Levelt, Mulders and
Schreiber~\cite{levelt} pointed 
out that semi-inclusive DIS could also be used to
study the flavor dependence of sea quarks. In particular,
\begin{eqnarray}
{\bar d(x) - \bar u(x) \over u(x) - d(x)} = {J(z)[1 - r(x,z)] - [1 + r(x,z)]
\over J(z)[1 - r(x,z)] + [1 + r(x,z)]},
\label{eq:3.5.4}
\end{eqnarray}
where
\begin{eqnarray}
r(x,z) = {d\sigma_p^{\pi^-}(x,z)/dz - d\sigma_n^{\pi^-}(x,z)/dz \over 
d\sigma_p^{\pi^+}(x,z)/dz - d\sigma_n^{\pi^+}(x,z)/dz},~~ J(z) =
{3 \over 5}~ {1 + D_u^{\pi^-}(z)/D_u^{\pi^+}(z) \over 
1 - D_u^{\pi^-}(z)/D_u^{\pi^+}(z)}.
\label{eq:3.5.5}
\end{eqnarray}
Unlike the situation for $d_v(x)/u_v(x)$ which is completely independent of
the fragmentation functions, Equation~\ref{eq:3.5.4} shows that
fragmentation functions are needed to extract the values of 
$\bar d(x) - \bar u(x)$. 

The HERMES collaboration~\cite{hermes98} at
DESY recently reported their measurements of charged hadrons produced in the
scattering of a 27.5 GeV positron beam on internal hydrogen, deuterium, and 
$^3$He target. The fragmentation functions $D_i^{\pi^\pm}(z)$
were extracted from the $^3$He data, while the hydrogen and deuterium
data allowed a determination of $r(x,z)$. The values of $(\bar d - \bar u)/
(u - d)$ show no $z$ dependence and are positive
over the region $0.02 < x < 0.3$, showing clearly an excess of $\bar d$ over
$\bar u$. Using the GRV94 LO~\cite{grv94} 
parametrization of $u(x) - d(x)$, the HERMES
collaboration obtained $\bar d(x) - \bar u(x)$  as shown 
in Fig.~\ref{fig:3.4.3}. The integral of $\bar d - \bar u$ over the
measured $x$ region gives 
$\int_{0.02}^{0.3} \left[\bar d(x) - \bar u(x)\right] dx = 
0.107 \pm 0.021(stat) \pm 0.017(syst)$. The total integral over all
$x$ is extrapolated to be $0.16 \pm 0.03$, consistent with the result from
NMC. 
%The $\bar d - \bar u$ values obtained from E866 are also shown in
%Figure~\ref{fig:3.5.1}. 
It is gratifying that the results from E866 and
HERMES are in rather good agreement, even though these two 
experiments use very different methods and cover very different 
$Q^2$ values ($\langle Q^2\rangle = 54$ GeV$^2$ in E866 
and $\langle Q^2\rangle = 2.3$ GeV$^2$ in HERMES).

It should be mentioned that semi-inclusive DIS could 
be extended~\cite{frank89,close91} to
situations involving polarized lepton beam and polarized targets in order
to study the flavor dependence of the spin-dependent structure functions.
Both the Spin Muon Collaboration (SMC)~\cite{smc98} 
and the HERMES Collaboration~\cite{hermes99} have
reported the polarized valence quark distributions, $\Delta u_v(x)$ and
$\Delta d_v(x)$, and the non-strange sea-quark polarization, 
$\Delta \bar q(x)$.

\subsection{\it Impact on the Parton Distribution Functions}
\begin{figure}[tb]
\begin{center}
\begin{minipage}[t]{11 cm}
\epsfig{file=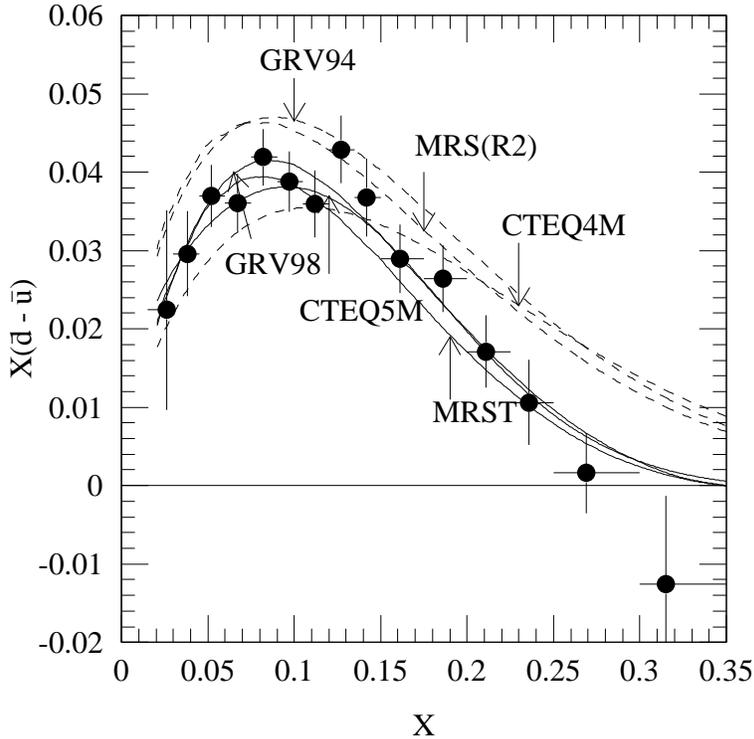,scale=1.2}
\end{minipage}
\begin{minipage}[t]{16.5 cm}
\caption{Comparison between the $x(\bar d - \bar u)$ results from E866
with the parametrizations of various parton distribution functions.
The dashed (solid) curves correspond to PDFs before (after) the E866
results were obtained.}
\label{fig:3.6.1}
\end{minipage}
\end{center}
\end{figure}

After the evidence for a flavor asymmetric sea was reported by the NMC
and NA51, several groups~\cite{cteq,mrs,grv94} performing global analysis of 
parton distribution functions all required $\bar d$ to be different
from $\bar u$. The NMC result constrained
the integral of $\bar d - \bar u$ to be $0.149 \pm 0.039$, while the NA51
result requires $\bar d / \bar u$ to be $1.96 \pm 0.13$ at $x = 0.18$. Clearly,
the $x$ dependences of $\bar d - \bar u$ and $\bar d/ \bar u$ were
undetermined. Figures~\ref{fig:3.4.2} and ~\ref{fig:3.6.1} compare
the E866 measurements of $\bar d/ \bar u$ and $x(\bar d - \bar u)$ 
with the parametrizations of the MRS(R2)~\cite{mrs} and CTEQ4M~\cite{cteq}, 
two of the most frequently used PDF's prior to E866's measurement.

Recently, several PDF groups published new parametrizations taking
into account of new experimental information including the E866 data.
The parametrization of the $x$ dependence of $\bar d - \bar u$ is now
strongly constrained by the E866 and HERMES data. In particular,
$\bar d(x) - \bar u(x)$ or $\bar d(x)/\bar u(x)$ are parametrized
as follows;

\noindent MRST~\cite{mrst}:
\begin{eqnarray}
\bar d(x) - \bar u(x) = 1.29 x^{0.183} (1-x)^{9.808} (1+9.987x-33.34x^2),
~Q_0^2 = 1~GeV^2,  \nonumber
\end{eqnarray}
GRV98~\cite{grv98}:
\begin{eqnarray}
\bar d(x) - \bar u(x) = 0.20 x^{-0.57} (1-x)^{12.4} (1-13.3x^{0.5}+60.0x),
~Q_0^2 = 0.4~GeV^2, \nonumber
\end{eqnarray}
CTEQ5M~\cite{cteq5, lai99a}:
\begin{eqnarray}
\bar d(x) / \bar u(x) = 1 - 1.095 x + 3.159 x^{0.5}/[1+(x-0.188)^2/0.01346],
~Q_0^2 = 1~GeV^2.
\label{eq:3.6.1}
\end{eqnarray}
As shown in Fig.~\ref{fig:3.4.2}, these new 
parametrizations give significantly different shape for $\bar d/ \bar u$
at $x > 0.15$ compared to previous parametrizations.
Table~\ref{tab:3.6.1} also lists the values of $\bar d - \bar u$ integral
from various various experiments and from recent PDF's.

\begin{table}[tbh]
\caption {Values of the integral $\int_0^1 (\bar d(x) - \bar u(x)) dx$ 
from various experiments and parton distribution functions.}
\begin{center}
\begin{tabular}{|c|c|}
\hline
Experiment/PDF & Integral \\
\hline\hline
NMC & $0.148 \pm 0.039$ \\
\hline
E866 & $0.118 \pm 0.012$ \\
\hline
HERMES & $0.16 \pm 0.03$ \\
\hline
CTEQ4M & 0.108 \\
\hline
MRS(R2) & 0.162 \\
\hline
GRV94 & 0.163 \\
\hline
CTEQ5M & 0.124 \\
\hline
MRST & 0.102 \\
\hline
GRV98 & 0.126 \\
\hline
\end{tabular}
\end{center}
\label{tab:3.6.1} 
\end{table}

\begin{figure}[tb]
\begin{center}
\begin{minipage}[t]{11 cm}
\epsfig{file=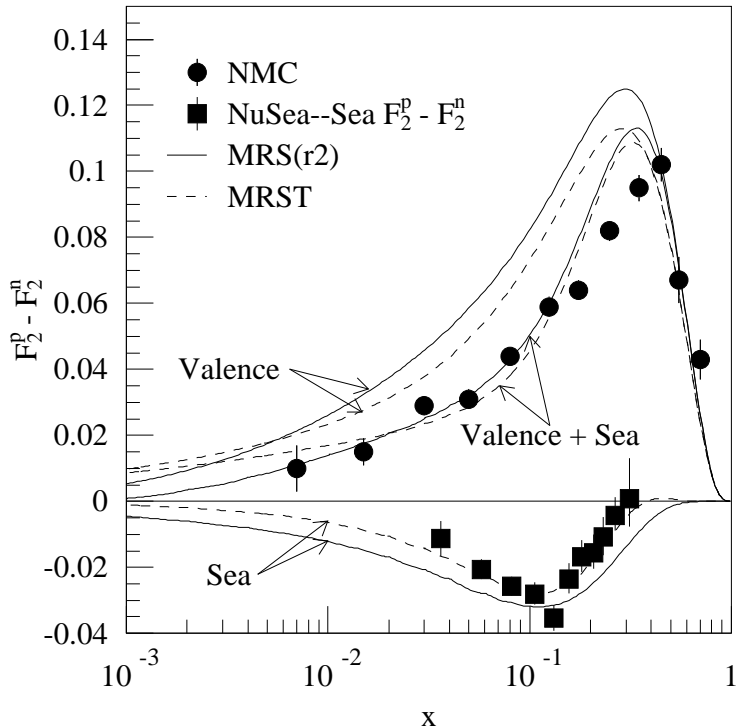,scale=1.2}
\end{minipage}
\begin{minipage}[t]{16.5 cm}
\caption{$F^p_2 - F^n_2$ as measured by NMC at $Q^2$ = 4 GeV$^2$ compared with
predictions based on the MRS(R2) and MRST parametrizations. Also
shown are the E866/NuSea results, rescaled to $Q^2$ = 4 GeV$^2$, for
the sea-quark contribution to $F^p_2
- F^n_2$. For each prediction, the top (bottom) curve is the valence
(sea) contribution and the middle curve is the sum of the two.}
\label{fig:3.6.3}
\end{minipage}
\end{center}
\end{figure}

It is interesting to note that the E866 data
also affect the parametrization of the valence-quark distributions.
Figure~\ref{fig:3.6.3} shows the NMC data for $F_2^p
- F_2^n$ at $Q^2$ = 4 GeV$^2$, together with the fits of MRS(R2) and
MRST. It is instructive to
decompose $F_2^p(x) - F_2^n(x)$ into contributions from 
valence and sea quarks:
\begin{eqnarray}
F_2^p(x) -F_2^n(x) = {1 \over 3} x \left[u_v(x) - d_v(x)\right] + {2 \over
3} x \left[\bar u(x) - \bar d(x)\right].
\label{eq:3.6.2}
\end{eqnarray}
As shown in Fig.~\ref{fig:3.6.3}, the E866 data
provide a direct determination of the sea-quark contribution to $F_2^p
- F_2^n$. In order to preserve the fit to $F_2^p - F_2^n$, the MRST's
parametrization for the valence-quark distributions, $u_v - d_v$,
is significantly lowered in the region $x > 0.01$. Indeed, one of the major new
features of MRST is that $d_v$ is now significantly higher than before at
$x > 0.01$. Although the authors of MRST attribute this to the new 
$W$-asymmetry data from CDF~\cite{cdf98} and 
the new NMC results on $F_2^d/F_2^p$~\cite{nmc97}, it 
appears that the new information on $\bar d(x) - \bar u(x)$ has a direct
impact on the valence-quark distributions too.

\begin{figure}[tb]
\begin{center}
\begin{minipage}[t]{11 cm}
\epsfig{file=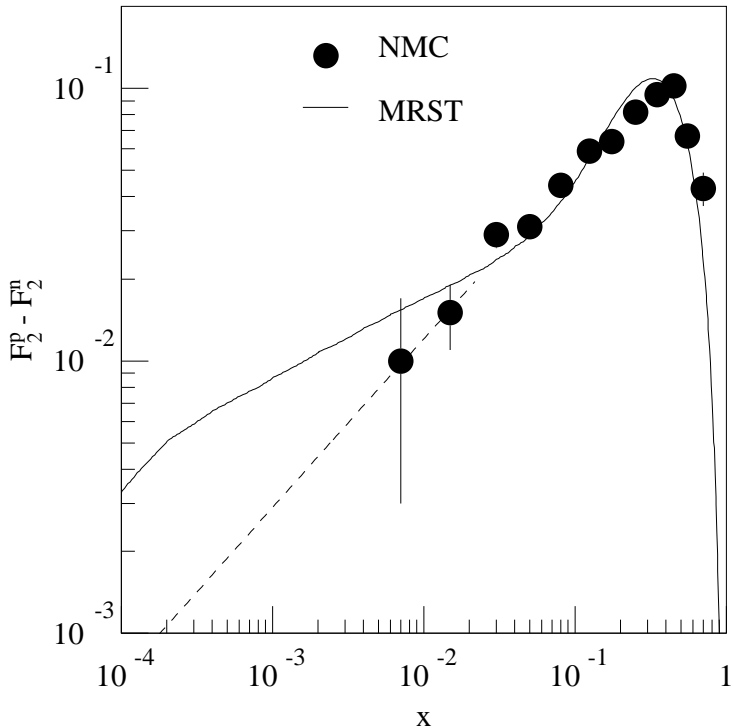,scale=1.2}
\end{minipage}
\begin{minipage}[t]{16.5 cm}
\caption{$F^p_2 - F^n_2$ as measured by NMC at $Q^2$ = 4 GeV$^2$ compared with
parametrization of MRST. The dashed curve corresponds to $0.21 x^{0.62}$,
a parametrization assumed by the NMC for the unmeasured small-$x$ region
when the Gottfried integral was evaluated.}
\label{fig:3.6.4}
\end{minipage}
\end{center}
\end{figure}

Another implication of the E866 data is on the behavior of
$F_2^p - F_2^n$ at small $x$. In order to satisfy the constraint
$\int_0^1 [u_v(x) - d_v(x)] dx = 1$, the MRST values of $u_v(x) - d_v(x)$
at $x < 0.01$ are now much larger than in MRS(R2), since
$u_v(x) -d_v(x)$ at $x > 0.01$ are smaller than before. 
As a consequence, $F_2^p - F_2^n$
is increased at small $x$ and MRST predicts a large contribution to the
Gottfried integral from the small-$x$ ($x < 0.004$) region, as shown in
Fig.~\ref{fig:3.6.4}. If the MRST
parametrization for $F_2^p - F_2^n$ at $x < 0.004$ were used, NMC
would have deduced a value of 0.252 for
the Gottfried integral, which would imply a value of 0.122 for 
the $\bar d - \bar u$ integral. This would bring excellent agreement 
between the E866 and the NMC results on the $\bar d - \bar u$ integral.

\subsection{\it Comparison Between Various Measurements}
Are the various measurements of sea quark flavor asymmetry consistent among 
them? In particular,
is the E866 result consistent with the earlier E772 and the NA51
DY experiments, and with the HERMES and NMC DIS measurements? To address 
this question, we first compare E866 with E772, both of which are 
DY experiments using 800 GeV proton beam with essentially the
same spectrometer. Figs.~\ref{fig:3.2.1} and~\ref{fig:3.2.2} show that
the MRST parton distributions, which determined $\bar d - \bar u$ based on 
the E866 data, can also describe the E772 data very well, and we conclude that
the E866 and E772 results are consistent.

Although both NA51 and E866 measured $\sigma (p+d)/ 2 \sigma (p+p)$ to extract
the values of $\bar d/ \bar u$, some notable differences exist. As mentioned
earlier, NA51 measured the ratio at a single value of $x_2$ ($x_2=0.18)$ near
$x_F \approx 0$ using a 450 GeV proton beam, while E866 used an 800 GeV
proton beam to cover a broader range of $x_2$ at $x_F > 0$.
It is instructive to compare the NA51 result at $x_2 = 0.18$ with
the E866 data at $x_2 = 0.182$. Table~\ref{tab:3.7.1} lists the
kinematic variables and physics quantities derived from these two 
data points. It is interesting to note that the values of 
$\sigma (p+d)/ 2 \sigma (p+p)$ at $x_2 = 0.18$ are actually 
very similar for NA51 and E866, even though the derived values 
for $\bar d/ \bar u$ differ significantly. This reflects the difference
in $x_F$ for both experiments, making the values of $\bar d/ \bar u$
extracted from $\sigma (p+d)/ 2 \sigma (p+p)$ different. The other difference
is $Q^2$, being a factor of 3.6 higher for E866. Using MRST and CTEQ5M 
to estimate
the $Q^2$ dependence of $\bar d/ \bar u$, we find that the NA51 value
of $\bar d/ \bar u$ is reduced by $\approx$ 3 \% going 
from $Q^2$ = 27.2 GeV$^2$ to $Q^2$ = 98.0 GeV$^2$. This brings slightly better
agreement bewteen NA51 and E866.

\begin{table}[tbh]
\caption {Comparison between the 450 GeV NA51 result and the 800 GeV 
E866 data point~\cite{e866} near $x_2 = 0.18$.}
\begin{center}
\begin{tabular}{|c|c|c|c|c|c|}
\hline
& $\langle x_2\rangle$ & $\langle x_F\rangle$ & 
$\langle M_{\mu \mu}\rangle$(GeV) & $\sigma^{pd}/ 2 \sigma^{pp}$ & 
$\overline d/ \bar u$ \\
\hline\hline
NA51 & 0.18 & 0.0 & 5.2 & 1.099 $\pm$ 0.039 & 1.96 $\pm$ 0.246 \\
\hline
E866 & 0.182 & 0.192 & 9.9 & 1.091 $\pm$ 0.044 & 1.41 $\pm$ 0.146 \\
\hline
\end{tabular}
\end{center}
\label{tab:3.7.1} 
\end{table}

The methods used by HERMES and E866 to determine $\bar d - \bar u$ are
different, and it is reassuring that the results came out
alike, as shown in Fig.~\ref{fig:3.4.3}. The $\bar d - \bar u$ values
from HERMES are in general somewhat larger than those of E866. 
At a relatively low mean $Q^2$ of 2.3 GeV$^2$, the HERMES experiment
could be subject to high-twist 
effects~\cite{ball94}. Additional data from HERMES
are expected to improve the statistical accuracy.

The comparison between E866 and NMC in terms of the integral of $\bar d -
\bar u$ has been discussed earlier. A possible origin for the apparent
differences of the integral was also discussed in Section 3.6.

\section{Origins of the $\bar d/ \bar u$ Asymmetry}

The earliest DIS experiments indicated that the Gottfried
integral was less than 1/3, leading to speculation regarding the origin
of this reduction. Field and Feynman suggested~\cite{field77} 
that it could be due to Pauli blocking in so far as $u \bar u$ pairs 
would be suppressed relative to $d \bar d$ pairs
because of the presence of two $u$-quarks in proton as compared to a single
$d$-quark.  Ross and Sachrajda~\cite{ross79} 
questioned that this effect would be
appreciable because of the large phase-space available to the created
$q \bar q$ pairs. They also showed that perturbative QCD would not 
produce a $\bar d$, $\bar u$ asymmetry. Steffens and Thomas~\cite{stef97} 
recently looked into this issue,
explicitly examining the consequences of  Pauli blocking. They
similarly concluded that the blocking effects were small, 
particularly when the antiquark is in a virtual meson.

The small $d, u$ mass difference (actually, $m_d > m_u$) of~ 2 to 4 MeV
compared to the nucleon confinement scale of 200 MeV does not permit any
appreciable difference in their relative production by gluons. At any rate,
one observes a surplus of $\bar d$ which is the heavier of the two species.
As pointed out above, blocking effects arising from the Pauli exclusion
principle should also have little effect. Thus another, presumably
non-perturbative, mechanism must be found to account for the large measured
$\bar d, \bar u$ asymmetry.

As many of the non-perturbative approaches to explain the asymmetry involve
the use of isovector mesons (particularly pions), we present some of the 
motivation for believing that pion field is intrinsic to the nucleon's
makeup.

\subsection{\it The Nucleon's Mesonic Field in Strong and Electroweak
Process}

The requirement of the pion as intrinsic to the nucleon has been evident to
most who have addressed the nucleon's interactions. To address the
nucleon's partonic structure as totally independent of its interactions 
seems naive indeed. However, the success of the constituent quark model 
in simply and directly explicating nucleon properties has focused 
attention on the quark structure of the nucleon to the point that
the role of pions was obscured. The magnetic moments of baryons are 
well characterized by constituent quarks, and pions are now known to 
have little effect on the result~\cite{dicus92}. In its most
extreme form there was a hope to start with three constituent quarks at
some low $Q^2$ scale and simply employ QCD evolution to generate 
nucleon's observed partonic distributions. We now know this is most 
unlikely if not impossible, as non-singlet quantities such as $g_A$,
the $\bar u$, $\bar d$ asymmetry, are of a non-perturbative origin and 
must be inserted ab initio if the nucleon's parton distributions are 
to be properly characterized. In the following we list properties of the 
nucleon that require it to possess a meson cloud.

\begin{figure}[tb]
\begin{center}
\begin{minipage}[t]{7 cm}
\epsfig{file=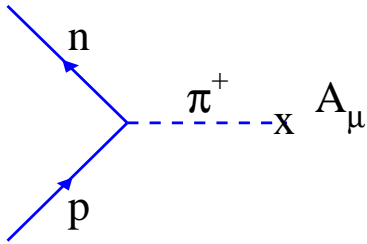,scale=1.4}
\end{minipage}
\begin{minipage}[t]{16.5 cm}
\caption{Diagram showing pion's role in nucleon beta decay.}
\label{fig:4.1.1}
\end{minipage}
\end{center}
\end{figure}

\begin{itemize}

\item The nucleon's strong interactions, particularly the long-range
part of the nucleon-nucleon interaction have been characterized via 
meson exchange. The development of a low energy nucleon-nucleon potential 
has gone on for many years~\cite{yukawa35,gammel57} with the long-range 
part in particular requiring a dominant role for pion exchange. Attempts 
to generate this interaction from QCD-inspired 
models~\cite{myhrer88,glozman00} have not met with quantitative
success~\cite{bartz01} so need for meson-exchange to account for the 
medium- and long-range parts of the nucleon-nucleon interaction appears 
beyond doubt.

\item The requirement that the nucleon axial current be partially
conserved (PCAC) requires the pion to be an active participant in the
nucleon. As the pion is the axial charge, it has a dominant status in
PCAC. Figure~\ref{fig:4.1.1} is a diagram of the pion's role in nucleon
beta decay.
Employing PCAC, the Goldberger-Treiman~\cite{gt58} relation 
can easily be derived,
\begin{eqnarray}
g_A = {{F_\pi g_c} \over {M_p}}.
\label{eq:4.1.1}
\end{eqnarray}
\noindent In the above expression $F_\pi$ is the pion decay constant 
(92.42 $\pm$ 0.26 MeV), $g_c$ is the charged pion ($\pi n p$)
coupling constant ($4\pi (14.17\pm 0.2)^{1/2}$)~\cite{ericson00}
and $M_p$ the proton mass. This yields a value for $g_A$ that is 
(3.8 $\pm$ 2.5) \% too high, not inconsistent with the variance expected 
due to the breaking of chiral symmetry ($m_u, m_d, m_\pi \neq 0$).
Furthermore, the value of the induced pseudoscalar form factor, $g_p(Q^2)$,
is also directly dependent on the pionic field of the nucleon. PCAC yields 
a result~\cite{bernard94}
\begin{eqnarray}
g_p(Q_0^2 = -0.88 m_\mu^2) = 8.44 \pm 0.23,
\label{eq:4.1.2}
\end{eqnarray}
consistent, albeit with large errors, with the measured value,
\begin{eqnarray}
g_p(Q_0^2) = 8.7 \pm 2.9.
\label{eq:4.1.3}
\end{eqnarray}
There are of course many other examples where mesons, especially 
the pion are employed to account for the nucleon's properties. Indeed, 
QCD based chiral models have treated the pion as a basic degree of freedom 
from the earliest times~\cite{georgi84}. However, the consequences
of their effects were not incorporated into parton distributions until
the early 90's~\cite{botts93,martin93}.

\end{itemize}
                             
As is well known, the information on the nucleon's structure from
parity-conserving elastic electron scattering is contained in four form
factors. They are the Dirac ($F_1(Q^2)$) and Pauli ($F_2(Q^2)$)
form factors for the neutron and proton. Linear combinations of these 
form factors are often employed, for example an isoscalar ($F^p+F^n$)
and an isovector ($F^p-F^n$) combination as well as the Sachs~\cite{sachs62}
electric and magnetic form factors:
\begin{eqnarray}
G_E(Q^2) = F_1(Q^2) - {Q^2 \over 4 M^2} F_2(Q^2);~~~~~ G_M(Q^2) = F_1(Q^2) +
F_2 (Q^2).
\label{eq:4.1.4}                                                       
\end{eqnarray}                                      
The normalization of the form factors are
$F_1^p(0) = 1$, $F_1^n(0) = 0$, $F_2^p(0) = 1.793$, and $F_2^n(0) =
-1.913$. The Sachs form factors are just the Fourier transforms of the 
charge and magnetization distributions in the Breit frame.

The large body of electron scattering data spanning $0 < Q^2 < 30$ GeV$^2$
can be analyzed in a largely model independent fashion using dispersion 
relations. They fit the data rather well but are not very revealing of 
the underlying structure of the nucleon. The electric form factor of 
the neutron on the other hand could be most revealing regarding the 
presence of a $p \pi^-$ component in the neutron wave function. There 
are two difficulties to be faced. First, the presence of the Foldy term 
in $G_E^n(Q^2)$ tends to obscure and complicate the interpretation of 
the neutron charge distribution~\cite{isgur99}. Secondly, there are 
QCD hyperfine interactions~\cite{isgur99,carlitz76,isgur81} that produce a
negative neutron charge radius similar to the $p \pi^-$ component. Thus it 
will take extensive additional theoretical and experimental work to clarify 
this important issue in electron-nucleon scattering. However,
some promising steps are being taken~\cite{wally99}.

\begin{figure}[tb]
\begin{center}
\begin{minipage}[t]{9 cm}
\epsfig{file=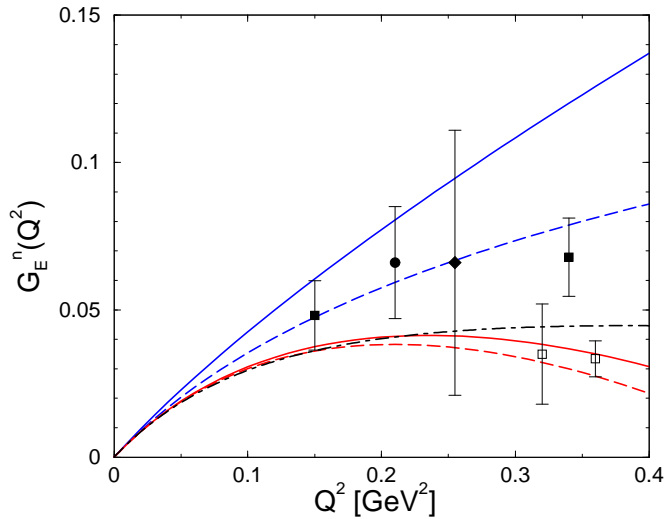,scale=0.5}
\end{minipage}
\begin{minipage}[t]{16.5 cm}
\caption{Comparison between data and calculations~\cite{kubis01} 
using relativistic baryon chiral perturbation theory for $G_E^n(Q^2)$.
The dashed curves are the 
results of calculation without vector mesons, the upper dashed curve 
being to 3rd order while the lower is to 4th order. The solid curves 
are the results including vector mesons where the upper curve is to 
3rd order while the lower is to 4th order. The dot-dashed curve is 
the result of the dispersion theoretical analysis.}
\label{fig:4.1.2}
\end{minipage}
\end{center}
\end{figure}

An area where appreciable recent progress has been made is in the analysis
of the nucleon's electromagnetic form factors in terms of relativistic baryon
chiral perturbation theory~\cite{kubis01}. This approach is the effective 
theory of QCD at low energy, which limits its range of applicability 
to $\sim Q^2 < 0.5 $GeV$^2$. The vector mesons are brought in
without the introduction of additional parameters by using the
results of theoretical dispersion analysis of the form factors. 
The neutron charge form factor is little changed by the inclusion of vector 
mesons, while they have a beneficial effect on the three other nucleon form 
factors. Figure~\ref{fig:4.1.2} shows the level of agreement obtained 
in $G_E^n(Q^2)$ applying relativistic baryon chiral perturbation theory as 
shown in Ref.~\cite{kubis01}. While the fits are impressive it still 
remains illusive as to what is added to our picture of the nucleon. The
pion field determines most of the properties of the form factors.
For example, the leading terms for the radii of the isovector charge and
magnetic form factors~\cite{wally99} are
\begin{eqnarray}
\langle r_1^v \rangle ^2 = {10 g_A^2 + 2 \over
(4\pi F_\pi)^2} ln {M_\pi \over m_N};~~~~~~
\langle r_2^v \rangle ^2 = {g_A^2 m_N \over
8\pi F_\pi^2 \kappa_v M_\pi},
\label{eq:4.1.6}
\end{eqnarray}

\noindent where $\kappa_v$ is the nucleon anomalous 
isovector magnetic moment (3.706). Note that in the chiral limit 
($M_\pi \to 0$), both radii become infinite, a not surprising 
result for massless pions. Our main purpose is to show that the pion mass
is a critical parameter in the nucleon form factors in this chiral model
derived from QCD. The important role of pions in the structure of the nucleon
appears to us to be beyond question and the nucleon's parton distributions 
should reflect this.

\subsection{\it Meson-Cloud Models}
As the $\bar u, \bar d$ asymmetry cannot be generated via perturbative 
process, several non-perturbative models have been proposed that yield 
an asymmetry. The most straightforward of these models are those that 
attribute the asymmetry to the existence of a ``pion cloud" in the proton. 
The relevance of pion cloud for sea-quark distributions appears to have first
been made by Thomas in a publication~\cite{thomas83}
treating SU(3) symmetry breaking in
the nucleon sea. Sullivan~\cite{sullivan}
had earlier shown that virtual meson-baryon
states directly contribute to the nucleon's structure function. A large
number of authors have contributed to calculating the asymmetry from
this perspective, so recent reviews~\cite{kumano98,speth98} should
be consulted for a complete list of contributions.

In the meson-cloud model, the virtual pion is emitted by the proton and the 
intermediate state is pion + baryon. More specifically, the proton is 
taken to a linear combination of a ``bare" proton plus pion-nucleon and 
pion-delta states, as below,
\begin{eqnarray}
|p\rangle \to \sqrt{1-a-b} |p_0\rangle + \sqrt{a} 
(-\sqrt{1 \over 3} |p_0\pi^0\rangle + \sqrt{2 \over 3} |n_0\pi^+\rangle) 
+ \sqrt{b} (\sqrt{1 \over 2} |\Delta^+_0
\pi^-\rangle - \sqrt{1 \over 3} |\Delta^+_0 \pi^0\rangle + \sqrt{1 \over
6} |\Delta^0_0 \pi^+\rangle).
\label{eq:4.2.1}
\end{eqnarray}
The subscript zeros on the virtual baryon states indicate that they are 
assumed to have symmetric seas, so the asymmetry in the antiquarks must be
generated from the pion valence distribution. The coefficients $a$ and $b$ 
are the fractions of the $\pi N$ and $\pi \Delta$ configurations, 
respectively, in the proton. These fractions can be calculated using the 
$\pi N N$ and $\pi N \Delta$ couplings, and form factors as taken from 
experiment. The asymmetry in the proton sea arises because of the 
dominance of $\pi^+$ among the virtual configurations. These
calculations, to be discussed in detail below, reproduce many aspects of
the data but suffer two problems. First, there is a strong dependence 
on the value used to cut off the integral over the form factor, and secondly,
any such calculation of $\bar d(x)/\bar u(x)$ is unreliable because the 
magnitude of the symmetric sea is unknown. Before launching into a more 
detailed presentation of these calculation, a few general observations 
can be made. It is instructive to examine some general properties of the 
$\pi N$, $\pi \Delta$ ansatz. A useful expression to consider is
\begin{eqnarray}
{\bar d_p(x) \over \bar u_p(x)} = {{5 \over 6} a f^N_\pi(x) +
{1 \over 3} b f^\Delta_\pi(x) + {1 \over 2} S(x) \over
{1 \over 6} a f^N_\pi(x) + 
{2 \over 3} b f^\Delta_\pi(x) + {1 \over 2} S(x)}. 
\label{eq:4.2.2}
\end{eqnarray}
$S(x)$ is the amount of symmetric sea and  $f^N_\pi(x) (f^\Delta_\pi(x))$
are functions that characterize the distributions in $x$ associated 
with the $\pi N$ ($\pi \Delta$) configurations. From this expression 
it is clear that the magnitude of the symmetric sea must be known 
if the ratio is to be predicted. The maximum value that the ratio can 
assume is 5, if there were no symmetric sea and only $\pi N$ configurations 
contribute. The minimum value of the ratio is $1/2$ which occurs in 
the absence of a symmetric sea and only $\pi \Delta$ configurations 
contributing. The ratio takes on a value 1 with a pure symmetric sea, 
or a symmetric sea with $a = b/2$ and $f^N_\pi = f^\Delta_\pi$. 
Pion models are much more effective 
in dealing with the integral isolating the contribution from
asymmetric sea (AS)
\begin{eqnarray}
I_{AS} \equiv \int_0^1 [\bar d_p(x) - \bar u_p(x)] dx = {1 \over 3} (2a-b),
\label{eq:4.2.3}
\end{eqnarray}
as there is no contribution from the symmetric sea in this case.

Many attempts~\cite{henley,kumano,signal,hwang,szczurek,koepf,wally98} 
have been made to calculate the flavor asymmetry due to
isovector mesons. Most start with the following convolution expressions:
\begin{eqnarray}
x\bar q_p(x,Q^2) = \sum_{MB} a^p_{MB} \int_x^1 dy~ f_{MB}(y)~ {x \over y}~
\bar q_M({x \over y},Q^2),
\label{eq:4.2.5}
\end{eqnarray}
where
\begin{eqnarray}
f_{MB}(y) = {g^2_{MpB} \over 16\pi^2} y \int_{-\infty}^{t_{min}} dt 
{F(t,m_p,m_B) \over (t - m^2_M)^2} F^2_{MpB}(t,\Lambda),~~ 
%\nonumber \\
t_{min} = m^2_p y - m^2_B {y \over 1-y}.
\label{eq:4.2.6}
\end{eqnarray}
In the above expressions $x$ is the fraction of proton''s momentum carried by
the antiquark, and
$y$ is fraction carried by the meson ($M$). The meson-proton-baryon 
couplings are
characterized by coupling constants $g_{MpB}$, and form 
factors $F_{MpB}(t,\Lambda)$ where $\Lambda$ is a cutoff parameter. 
$F(t,m_p,m_B)$ is a kinematic factor depending
on whether $B$ is in the baryon octet or decuplet. 
As pions are the only mesons usually considered and the baryons are usually
restricted to nucleons and deltas, the coupling constants are well known
and the partonic structure of the pion, $\bar q_\pi (x,Q^2)$, is fixed 
by measurement of the
DY process using high energy pion beams. The only uncertainties are
the form factors $F_{\pi pN}(t)$ and $F_{\pi p \Delta}(t)$. One 
attempts to determine these form
factors by using~\cite{holt96,niko99} the measured yields from 
a variety of high energy hadronic reactions at small 
$p_T$ such as $p p \to n X$ and $p p \to \Delta^{++} X$.
Even though there is a sizable amount of available data, employing such a
procedure does not produce a precise result. The cutoff parameters used 
in the extracted 
form factors are of the order of 1 GeV  but the uncertainties in their values
produce factors of two in the predicted antiquark content of the nucleon due
to pions.

\begin{figure}[tb]
\begin{center}
\begin{minipage}[t]{11 cm}
\epsfig{file=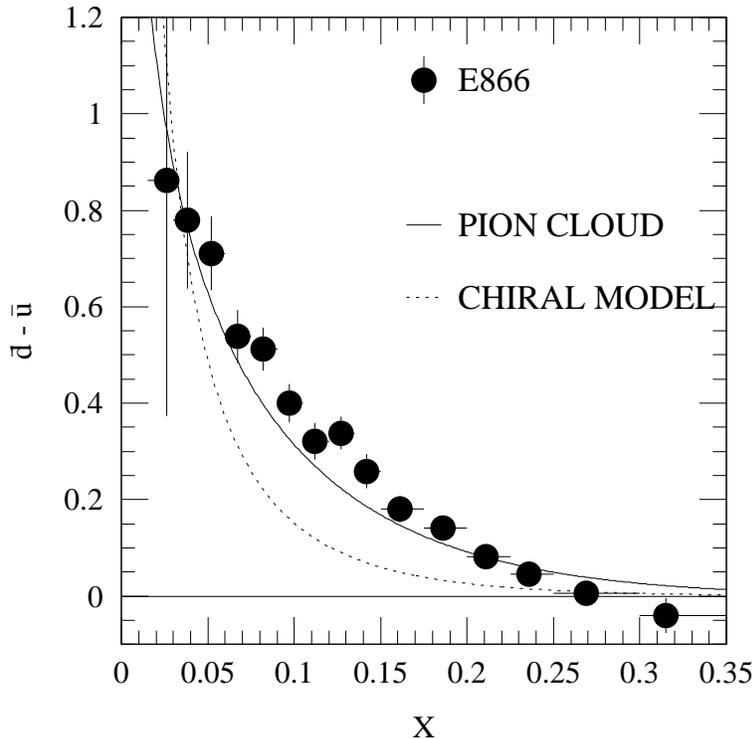,scale=1.2}
\end{minipage}
\begin{minipage}[t]{16.5 cm}
\caption{Comparison of the E866~\cite{e866} $\bar d - \bar u$ results at 
$Q^2$ = 54 GeV$^2$/c$^2$ with the predictions of pion-cloud and chiral models 
as described in the text.}
\label{fig:4.2.1}
\end{minipage}
\end{center}
\end{figure}

Even though calculated value of the integral of $\bar d(x) - \bar u(x)$
is often in agreement with
experiment, it is more difficult to achieve a quantitative fit to the
measured $x$ dependence of the difference~\cite{peng98}. 
Figure~\ref{fig:4.2.1} compares $\bar d(x) - \bar u(x)$ from E866
with a pion-cloud model calculation, following the procedure detailed
by Kumano~\cite{kumano}. A dipole form, with $\Lambda = 1.0$ GeV for 
the $\pi N N$ form factor and $\Lambda = 0.8$ GeV for 
the $\pi N \Delta$ form factor,
was used. Calculations of $\bar d(x)/ \bar u(x)$  
are even more unsuccessful, as
knowledge of the $x$ dependence of the symmetric sea is required in this
instance~\cite{peng98}. 

\begin{figure}[tb]
\begin{center}
\begin{minipage}[t]{11 cm}
\epsfig{file=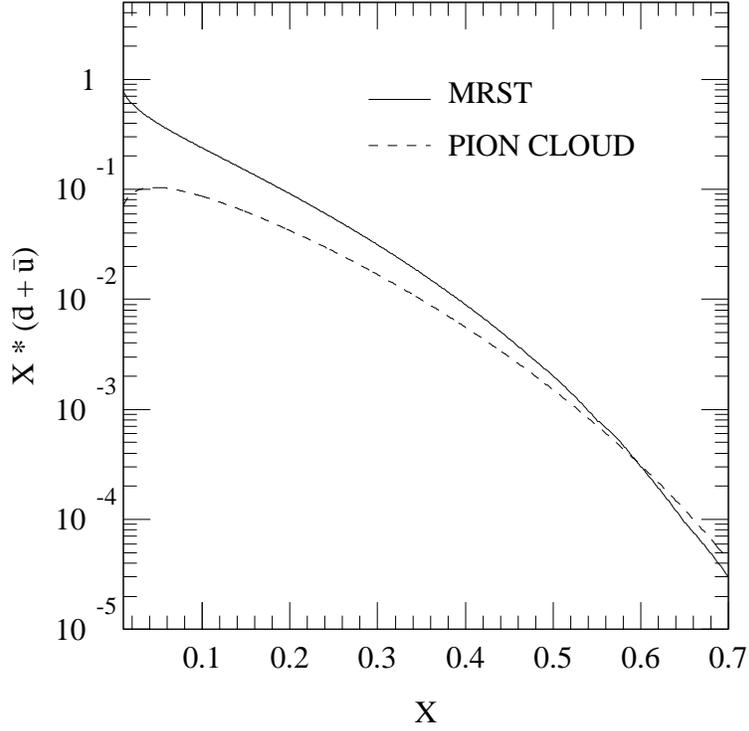,scale=1.2}
\end{minipage}
\begin{minipage}[t]{16.5 cm}
\caption{Comaprison of $x (\bar d + \bar u)$ obtained from the valence
quarks in the pion cloud with
the parametrization of the MRST parton distribution functions.}
\label{fig:4.2.2}
\end{minipage}
\end{center}
\end{figure}

It is instructive to compare the pion-model prediction with
the current PDF parametrization of $x(\bar d + \bar u)$.
Figure~\ref{fig:4.2.2} shows that at small $x$ ($x < 0.1)$ 
the valence quarks in the
pion cloud account for less than $1/3$ of the $\bar d + \bar u$ content in the
proton. In contrast, at large $x$ ($x > 0.5$) the pion model would 
attribute all of $\bar d + \bar u$ to the pion cloud.

\begin{figure}[tb]
\begin{center}
\begin{minipage}[t]{9.5 cm}
\epsfig{file=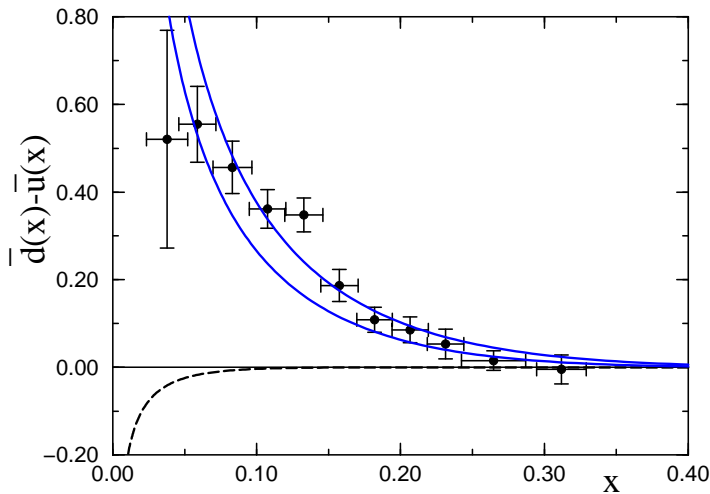,scale=0.7}
\end{minipage}
\begin{minipage}[t]{16.5 cm}
\caption{Flavor asymmetry at $Q^2 = 54$ GeV$^2$. Experimental data are
from E866~\cite{e866}. The solid curves are from Nikolaev et al.~\cite{niko99}
for the $\pi N$ Fock state with Gaussian form factors $R_G^2 = 1$GeV$^2$ (upper)
and $R_G^2 = 1.5$GeV$^2$ (lower). The dashed line shows the contribution of the
$\pi \Delta$ Fock state for a Gaussian form factor with $R_G^2 = 2$ GeV$^2$.}
\label{fig:4.2.3}
\end{minipage}
\end{center}
\end{figure}

Determining the appropriate cutoff is difficult because the high energy
hadronic reactions ($p p \to \pi X$, $p p \to n X$, $p p \to \Delta^{++} X$)
used to fix the form factors all have experimental backgrounds, and 
extracting quantitative results from such reactions is difficult. 
Nikolaev et al.~\cite{niko99} carefully investigated these
reactions by including constraints from the Regge behavior of
various mesons on the total photo-absorption cross section. They find that
the contribution of $\rho$ and $a_2$ Reggeons to the proton structure 
function is negligble. They also find the relative contribution of 
$\pi \Delta$ to be much smaller than previous analysis. For example 
$a = 0.105, b = 0.015$ for cutoffs of $R^2_G = 1.5 $GeV$^{-2}$ 
and $R^2_G = 2$ GeV$^{-2}$ for 
the $\pi N$ and $\pi \Delta$ respectively. Their fits to $\bar d_p(x) -
\bar u_p(x)$ are shown in Fig.~\ref{fig:4.2.3}.

In a recent publication~\cite{thomas00} Thomas et al. appear to have 
avoided the form factor issue by having recourse to Chiral Perturbation 
Theory~\cite{chiral}. In this case the theory is implemented with pions, 
nucleons and deltas as the degrees of freedom. They find for the 0'th
moment of the $\bar d_p(x) - \bar u_p(x)$ due to the $\pi N$ loop to be
\begin{eqnarray}
I^{LNA}_{AS}|_{\pi N} \equiv \int_0^1 dx [\bar d(x) - \bar u(x)]
= {2g^2_A \over (4\pi f_\pi)^2} m^2_\pi ln {m^2_\pi \over \mu^2},
\label{eq:4.2.7}                      
\end{eqnarray}
where $g_A$ is the axial charge of the nucleon, $f_\pi$ the pion decay 
constant and $\mu$ a mass parameter. They stress that this leading 
non analytic (LNA) behavior originates from one-pion loop and hence 
is robust. General expressions are derived for the n'th order terms 
which are suppressed by powers of $(m_\pi/M_p)^n$.
Using Eqs. (2), (5), (6) and (8) from Ref.~\cite{thomas00} one obtains 
the ratio of the contribution of  $\pi \Delta$ to that of $\pi N$ to be
\begin{eqnarray}
{I^{LNA}_{AS}|_{\pi \Delta} \over I^{LNA}_{AS}|_{\pi N}}
= - {3 \over 75} {(M_\Delta + M_N)^2 \over M^2_\Delta} = -0.12.
\label{eq:4.2.8}                      
\end{eqnarray}
Evaluating the net result using $\mu = 4 \pi f_\pi$ one finds
\begin{eqnarray}
I^{LNA}_{AS} = 0.154
\label{eq:4.2.9}                      
\end{eqnarray}
close to the E866 value of 0.118$\pm$0.012.

The values for $I_{AS}$ that result from the various parametrizations 
of the $\pi N$ and  $\pi \Delta$ virtuality are all reasonable and not 
far from the experimentally measured value of 0.118$\pm$0.012. 
The values for $a$ and $b$ 
from various authors who have
employed the meson cloud model show
that a typical value of $b/a$ is $1/2$. This leads to 
$a=0.24$, $b=0.12$ to satisfy the observed flavor asymmetry, resulting 
in the probability of finding a pion in a nucleon (Eq.~\ref{eq:4.2.1})
of $a+b = 0.36$.

The $\pi N$, $\pi \Delta$ model also allows one to evaluate the effect 
of virtual meson emission on the spin carried by quarks. The pion is 
emitted from the proton in a $p$-wave and therefore reduces the spin 
projection of the virtual baryon, thus reducing the spin projection 
carried by quarks. In this picture there is no spin on the antiquarks 
and the reduced spin in the baryon sector is carried off in the orbital 
motion of the pions. The expressions below for the spin on the quarks 
assumes an SU(6) wave function
for the baryons. 
\begin{eqnarray}
\Delta u + \Delta d = 1 - {2 \over 3} (2a-b) = 1 - 2 I_{AS}
\label{eq:4.2.10}                      
\end{eqnarray}
\begin{eqnarray}
\Delta u - \Delta d = {5 \over 3} - {20 \over 27}(2a + b) + {32 \over 27}
\sqrt{2ab}
\label{eq:4.2.11}                      
\end{eqnarray}
Where
\begin{eqnarray}
\Delta q \equiv \int_0^1 [q(x) \uparrow + \bar q(x) \uparrow - q(x) \downarrow
- \bar q(x) \downarrow ] dx.
\label{eq:4.2.12}                      
\end{eqnarray}

Interference occurs in Eq.~\ref{eq:4.2.11} because the operator 
$\sigma_Zt_Z$ connects nucleon and delta states, while in Eq.~\ref{eq:4.2.10}
the operator $\sigma_Z$ cannot. Thus, virtual pion emission is
seen to reduce the spin carried by quarks. With no further assumptions we
have $\Delta u + \Delta d = 0.76 \pm 0.024$, which is appreciably smaller 
than one but still greater than the observed value of $0.40 \pm 0.02$. 
Employing a commonly used value of $b \simeq a / 2$, we obtain 
$\Delta u - \Delta d = 1.51 \pm 0.053$. 
Again virtual pion emission reduces $\Delta u - \Delta d \equiv g_A$
from its SU(6) value of 5/3 but not nearly to
the measured value of 1.256. Interestingly, the value extracted for
$g_A$ using the measured flavor asymmetry in this model is nearly identical
to that obtained by Weinberg~\cite{weinberg91} using chiral perturbtion 
theory and current algebra techniques.

\subsection{\it Chiral Models}
An alternative 
approach ~\cite{ehq92,li95,li97,weise98,szc96,song97,ohlsson99} also 
employing virtual pions to produce the $\bar d, \bar u$ asymmetry is 
usually referred to as chiral models~\cite{georgi84}. 
After the publication of the NMC result on the violation of the
Gottfried sum rule, Eichten, Hinchliffe and Quigg (EHQ)~\cite{ehq92}, 
following an earlier suggestion of Bjorken~\cite{bjorken}, 
published an interesting approach to explain the NMC result. They
used chiral perturbation theory, employing the Manohar and 
Georgi~\cite{georgi84} model in which the degrees of freedom are 
Goldstone Bosons (GB), and constituent quarks (i.e., $U$ and $D$ for 
up and down constituent quarks). The pion in this approach 
is emitted by a constituent quark. In this model most of the parameters 
are more or less prescribed (the mass of the constituent quark must 
be assigned). If $\alpha$ is the probability of a $U$ constituent quark 
becoming $\pi^+ D$, then to first order in $\alpha$, invoking isospin
invariance and the conservation of probability
\begin{eqnarray}
U \to (1-{3 \over 2}\alpha)U + \alpha \pi^+ D + {1 \over 2} \alpha \pi^0 U
\label{eq:4.3.1}
\end{eqnarray}
with a corresponding expression for the down constituent quark. As the
proton is $UUD$, to first order in $\alpha$ one has
\begin{eqnarray}
p \to 2U + D + {7\alpha \over 4} (u + \bar u) + 
{11\alpha \over 4} (d + \bar d).
\label{eq:4.3.2}
\end{eqnarray}
Thus the integrated asymmetry, $I_{AB}$, is equal to $\alpha$.
Note that in the absence of a symmetric sea the maximum value for
$\bar d(x) / \bar u(x)$ in this model is 11/7. Restricting themselves 
to only the pion the authors calculate a value of $\alpha = 0.083$.
This value 
is somewhat small, but within 50\% of the measured value.
However, it is very interesting that their calculation of $\alpha$, 
carried out in the context of chiral field theory~\cite{georgi84} 
generates a probability of finding a pion in a nucleon of $9\alpha /2 = 0.374$,
very similar to that found in the meson-cloud model.

The authors also investigated the consequences of  this model on the quark
spins. Using SU(6) wave function for the proton and assuming total
spin-flip of the constituent quark upon emitting a pion, they found for the
modified quark spins in the proton
\begin{eqnarray}
\Delta u + \Delta d = 1 -3\alpha = 1 - 3I_{AS}
\label{eq:4.3.3}
\end{eqnarray}
\begin{eqnarray}
\Delta u - \Delta d = {5 \over 3} - {5 \over 3} \alpha .
\label{eq:4.3.4}
\end{eqnarray}
Using $\alpha = 0.118 \pm 0.012$ from the E866
measured flavor asymmetry, $\Delta u + \Delta d = 0.65$ and
$\Delta u - \Delta d = 1.47$. Both are reduced from the SU(6) 
values but are not as small as the measured values of 0.40 and 1.256,
respectively. One believes that the SU(6) value is too large due to 
relativistic effects but the size of reduction is highly model dependent.

The differences that occur between the EHQ model and the meson-cloud model
employed above
are that the pion is emitted by a single constituent quark rather than by
the proton. EHQ have a single parameter $\alpha$ which corresponds to 
fixing $a/b$ at 5/4 in the meson-cloud model. We believe that this value is too 
small and does not easily produce the large values of  $\bar d(x) / \bar u(x)$
that are observed in the interval $0.1 < x < 0.2$. EHQ achieves greater 
suppression of the quark spin because they assume a total spin flip of 
the constituent quark upon emission of a pion.

Reference~\cite{ehq92} also examined the effect of extending the space 
of GBs beyond the pion. Treating the octet ($\pi, K, \eta$)
of GBs as degenerate with equal coupling to all members, 
the proton becomes to first order in $\alpha^\prime$,
\begin{eqnarray}
p \to 2U + D + 2\alpha^\prime(u + \bar u) + {8 \over 3} \alpha^\prime
(d + \bar d) + {10 \over 3} \alpha^\prime (s + \bar s).
\label{eq:4.3.5}
\end{eqnarray}
In this case the $\bar u,\bar d$ flavor asymmetry is diluted, with the maximum
value for $\bar d(x) / \bar u(x)$ becoming 1.333, and 
$I_{AS} = 2\alpha^\prime /3$.
Thus satisfying the E866 measured value for $I_{AS}$ requires $\alpha^\prime
= 0.177 \pm 0.017$. The effect on the proton spin is given by
\begin{eqnarray}
\Delta u - \Delta d = {5 \over 3} - {45\alpha^\prime \over 9}
\label{eq:4.3.6}
\end{eqnarray}
\begin{eqnarray}
\Delta u + \Delta d + \Delta s = 1 - {16 \alpha^\prime \over 3}.
\label{eq:4.3.7}
\end{eqnarray}
Both larger value for $\alpha^\prime$ and the larger number of mesons 
available to flip the quark spins conspire to produce a large effect 
on the spin values, $\Delta u - \Delta d = 0.978 \pm 0.066$ and
$\Delta u + \Delta d + \Delta s = 0.056 \pm 0.091$. It is clear 
that this approach produces too little flavor asymmetry and too much 
reduction of quark spin. In addition one can see from Eq.~\ref{eq:4.3.5}
that far too much mesonic sea has to be generated to produce the 
observed flavor asymmetry. Continuing in this vein, EHQ showed that 
the $\bar u, \bar d$ asymmetry vanishes in the case of an ideal U(3) 
nonet. This of course is not what is observed as the $\eta$ and 
$\eta^\prime$ contributions are suppressed because they 
are much more massive than the $\pi$.

This approach has been considerably extended by Cheng and Li~\cite{li95,li97,
li98,li98a}
They employ the full meson U(3) nonet but allow the symmetry to be broken
{$U(3) \to SU(3) \times U(1)$}. The coupling to the singlet $\eta^\prime$
is unique. Originally they made an ad hoc assumption that the singlet 
coupling was the negative of the octet coupling. Subsequently they 
have presented strong support for this point of view using the coupling 
of quarks to instantons~\cite{li99,cheng99} and using the same physics that 
produces the large mass for the $\eta^\prime$. Defining the ratio of 
the coupling to the singlet to that of the octet as $f_1/f_8 \equiv \zeta$,
they found $\zeta = -2$. The ratio $\bar d / \bar u$ in terms of 
$\zeta$, due to the full nonet of GBs is
\begin{eqnarray}
{\bar d \over \bar u} = {{8 \over 3} + {1 \over 3} \zeta^2  \over
2 + {2 \over 3} \zeta + {1\over 3} \zeta^2},
\label{eq:4.3.8}
\end{eqnarray}
which gives 1 for $\zeta = 1$, and 4/3 for $\zeta = 0$ (the pure 
octet case). With $\zeta = -2$ the ratio is 2, an altogether 
reasonable value. However, in their
most recent tabulation~\cite{li98a} of results they choose $\zeta = -1$
and generate the results shown in the third column of Table~\ref{tab:4.3.1}.
In order to account for expected suppression factors arising 
from mass differences, they included a weighting factor, 
$(\langle k^2\rangle +m^2_{GB})^{-1}$, for each GB amplitude. This result is shown
in column 4 and is seen to reduce the excessive contribution of strange
quarks. This description while apparently quite successful, has
the problem of producing a softer $x$ distribution 
for $\bar d(x) - \bar u(x)$ than is observed in E866. Figure~\ref{fig:4.2.1}
shows a chiral-quark model calculation~\cite{peng98} using the formulation
described in Ref.~\cite{szc96}. 
The predicted $\bar d(x) - \bar u(x)$ 
distribution is too soft because the GB is emitted by a constituent 
quark, which carries only 1/3 of the nucleon's momentum. This suggests 
that additional dynamics must be included in the chiral model if it is 
to produce agreement with the data.

\begin{table}[tbh]
\caption {Comparison of the measured parton flavor and spin structure of the
proton with the model of Cheng and Li~\cite{li98,li98a}. Some of the measured
values are slightly different from the current values. The value of $\langle
k^2 \rangle$ used is 350 MeV$^2$.}
\begin{center}
\begin{tabular}{|c|c|c|c|}
\hline
Quantity & Measured Value & SU3 & Broken SU3 \\
\hline\hline
$\overline d - \bar u$ & $0.147 \pm 0.026$ & 0.15 & 0.15 \\
\hline
$2\bar s/ (\bar u + \overline d)$ & $\sim 0.5$ & 1.86 & 0.6 \\
\hline 
$\Delta u$ & $0.82 \pm 0.02$ & 0.78 & 0.85 \\
\hline
$\Delta d$ & $-0.43 \pm 0.02$ & -0.33 & -0.40 \\
\hline
$\Delta s$ & $-0.10 \pm 0.02$ & -0.11 & -0.07 \\
\hline
$\Delta\Sigma$ & $0.29 \pm 0.06$ & 0.34 & 0.38 \\
\hline
$\Delta \bar u, \Delta \overline d$ & $0.01 \pm 0.07$ & 0.0 & 0.0 \\
\hline
$g_A$ & $1.257 \pm 0.03$ & 1.12 & 1.25 \\
\hline
\end{tabular}
\end{center}
\label{tab:4.3.1} 
\end{table}

By way of further comment on Table~\ref{tab:4.3.1}, there is no reason 
to expect the ratio of $2 \bar s / (\bar d + \bar u)$ generated from
GBs alone should be 0.5, as this ratio depends sensitively on the contribution
from the symmetric sea. Also there is a lack of consistency in 
introducing a mass 
factor in the GB sector and ignoring the effect in the baryon sector. 
The baryons accompanying the GBs are $N, \Delta, \Lambda, \Sigma$
with mass splittings comparable to those occuring among the GB.

Continuing with approaches that explain the flavor asymmetry and spin
quenching by introducing the role of mesons, a novel approach has been
taken by Ball et al.~\cite{forte94a,forte94b} in which the mesons are 
introduced by including them in the $Q^2$ evolution of the quark 
structure function functions as a non-singlet non-perturbative component. 
The approach is described in detail in~\cite{forte94a,forte94b} 
and incorporates many attractive features. For example they
explicitly account for the mass dependence of the contribution from each
of the pseudoscalar mesons. Their approach generates a very strong scale
dependence to the value observed for the Gottfried Sum Rule.
Unfortunately, there is no relevant DIS data 
to verify their prediction.  Drell-Yan experiments which could 
equivalently measure $I_{AS}$ as a function of $Q^2$ are typically 
restricted to using lepton pairs above the $\Psi^\prime$ ($M_{\Psi^\prime}=
3.69$ GeV),
so have minimum $Q^2$ of $\sim$ 16 GeV$^2$ making it impossible to observe 
the predicted scale dependence.

\begin{figure}[tb]
\begin{center}
\begin{minipage}[t]{13 cm}
\epsfig{file=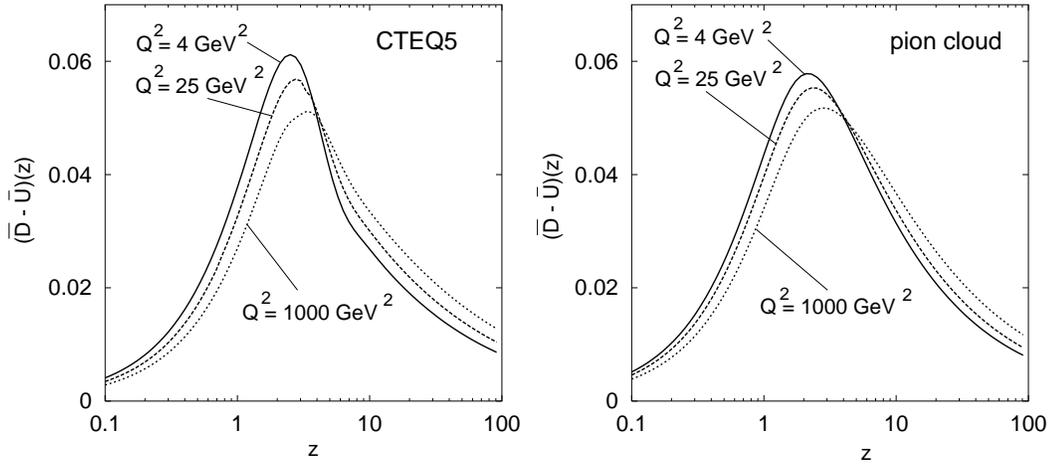,scale=0.7}
\end{minipage}
\begin{minipage}[t]{16.5 cm}
\caption{$(\bar D - \bar U) (z)$ in coordinate
space from CTEQ5 parton distributions and the pion cloud model~\cite{henley01}.}
\label{fig:4.3.1}
\end{minipage}
\end{center}
\end{figure}

A very nice paper recently published by Henley, Renk, and 
Weise~\cite{henley01} showed that the distribution in $x$ of
$\bar d(x) - \bar u(x)$ measured by E866 corresponds to that 
associated with a pion cloud. Their result is somehow
implicit in earlier work that showed the pion cloud model 
reproduced the $x$ distribution observed in experiment, however 
this work makes explicit that the spatial distribution involved is 
what one would expect for a pion cloud. The method employed uses the 
coordinate space representation developed earlier by Piller 
et al.~\cite{piller00} and V\"{a}nttinen et al.~\cite{piller98}.
They introduce a light-cone dimensionless space-time variable 
$z = y \cdot P$, where $P$ is the nucleon momentum. The light-cone 
distance is $y^+ \equiv t + y_3 = 2 z/M$, with
$M$ the nucleon mass. The dimensionless variable is 
conjugate to Bjorken $x$, and $z = 10$ corresponds to $y^+ = 4$ fm. 
The authors work with parton distribution functions such as CTEQ5 or 
MRST that accurately capture the
measured $\bar d(x) - \bar u(x)$ distributions. The transformation 
from the $x$ distribution is carried out via
\begin{eqnarray}
(\bar D - \bar U) (z,Q^2) \equiv \int_0^1 [\bar d(x,Q^2) -
\bar u(x,Q^2)] sin(zx) dx
\label{eq:4.3.9}
\end{eqnarray}
Figure~\ref{fig:4.3.1} shows how the $(\bar D - \bar U)(z)$ 
distributions appear when plotted against $z$. The peak in $z$ is at 
2.5 or about 1fm and the bulk of the distribution is between $z = 1$ and
10 corresponding to 0.4 to 4 fm, in line with
what one expects for a $p$-wave meson cloud.

\subsection{\it Other Approaches not Directly Including Mesons}
\subsubsection{\it Instanton Models}
Instantons have been known as theoretical constructs since the
seventies~\cite{bel75,hooft76,shu98}. 
They represent non-perturbative fluctuations of the gauge fields
that induce transitions between degenerate ground states of different
topology. In the case of QCD, the collision between a quark and an
instanton flips the helicity of the quark while creating a $q \bar q$
pair of different flavor. Thus, interaction between a $u$ quark and an 
instanton results in a $u$ quark of opposite helicity and either a $d \bar d$ 
or $s \bar s$ pair. 
Such a model has the possibility of accounting for both the flavor
asymmetry and the ``spin crisis"~\cite{forte89,forte91}. 
However, the model has proven difficult to exploit for
this purpose. There is only one case~\cite{inst93} of its being 
directly employed to explain these
anomalous effects. In the case of the $\bar d, \bar u$ flavor 
asymmetry, the authors of ref.~\cite{inst93} fit
the instanton parameters to reproduce the violation
of the GSR observed by NMC. 
The prediction~\cite{inst93} at large $x$, $\bar d(x) / \bar u(x) \to 4$, 
is grossly violated by experiment (see Fig.~\ref{fig:3.4.2}). 
Thus, it appears that while instantons have the
possibility for accounting for flavor and spin anomalies, the approach is not
yet sufficiently developed for a direct comparison. The final state created
via an instanton collision is quite similar to that created via the
emission of meson in the chiral model.

However, it must be pointed out that the instanton model contains elements
that strongly influence other descriptions. For example, as cited above 
in the work of Cheng et al.~\cite{li99,cheng99}, instantons provide the basis 
for generating a flavor asymmetry when the full GB nonet is employed. If 
the coupling to the $\eta^\prime$ is the same as the coupling to the GB 
octet, the sea is flavor symmetric. They originally made the ad hoc 
assumption that the $\eta^\prime$ coupling was equal in magnitude but 
opposite in sign to the coupling to the octet. This assumption has been 
justified~\cite{cheng99} by invoking instantons. Further, as will
be discussed below, the chiral quark soliton model which is developed 
using a vacuum sea of pions can equally well~\cite{diak96} be derived using 
the instanton model of the QCD vacuum. Thus it appears that the notion 
of instantons might be essential to the flavor asymmetry and that they 
play a crucial role in underpinning  what ever explanation is used.

\subsubsection{\it Lattice Gauge Approach}
\begin{figure}[tb]
\begin{center}
\begin{minipage}[t]{15 cm}
\epsfig{file=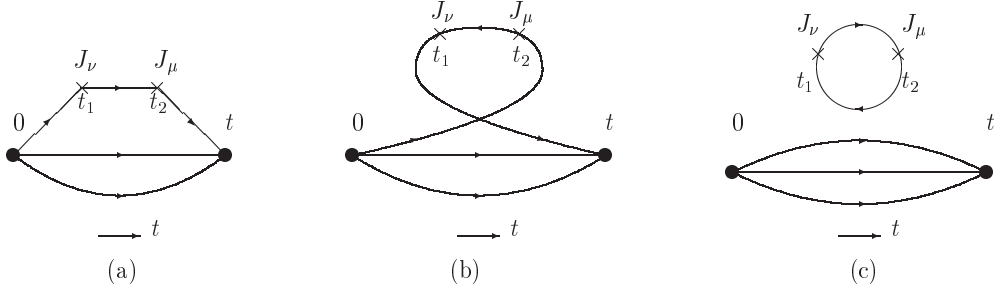,scale=0.8}
\end{minipage}
\begin{minipage}[t]{16.5 cm}
\caption{Euclidean path integral from Ref.~\cite{liu00} 
for evaluating the hadronic tensor
$W_{\nu\mu}$.
(a) and (b) are connected insertions and (c) is a
disconnected insertion.}
\label{fig:4.4.1}
\end{minipage}
\end{center}
\end{figure}

It would be extremely informative to explain the $\bar d, \bar u$
asymmetry in terms of lattice gauge calculations as it would provide 
some insight as how the evolution of the asymmetry takes place. 
Indeed, soon after the existance of the asymmetry was established 
K.F. Liu and his collaborators investigated~\cite{liu94,liu00}
the issue. Employing a path-integral formalism they established that 
the $\bar d, \bar u$ difference comes from the set of diagrams they 
termed connected diagrams 
(Fig.~\ref{fig:4.4.1} (a,b)).
The other sources of sea quarks are disconnected diagrams, shown in 
Fig.~\ref{fig:4.4.1}(c). The loops in the disconnected diagrams 
are generated by gluons, and can produce all flavors of sea quarks. 
These diagrams produce equal numbers of up and down sea quarks. 
The loops in the connected diagrams can only  produce up
and down sea quarks and presumably are the source of the up, down asymmetry.

This approach has the interesting feature that the strange quark sea is
created entirely from disconnected diagrams and hence the $s$ and $\bar s$
distributions are identical, in good agreement with the results from
neutrino DIS. It would also account for the near vanishing of the strange
vector form factors for the nucleon as extracted from parity violating
electron scattering. However it leaves unspecified the spin carried by
strange quarks.

\subsubsection{\it Chiral-Quark Soliton Model}

One of the interesting approaches to emerge over the past 5 years
is that of the chiral-quark soliton model~\cite{diak96,diak97,poby96,poby99,
waka98}. It uses the large $N_c$ limit of QCD, which becomes an effective 
theory of mesons with the baryons appearing as solitions. At low energy 
the effective dynamics is described by a chiral Lagrangian for the pion, 
which appears as the GB of the spontaneously broken chiral 
symmetry. The model employs a realistic effective interaction, containing 
all orders of derivatives of the pion field, defined by the integral over 
quark fields having dynamically generated mass and interacting with 
the pion field in a minimally chirally invariant way. It is valid over 
a range of momenta up to the inverse of the instanton size 
($1/\rho = 0.6$ GeV). The parton distribution obtained at this
low momentum scale can then be evolved to 
higher $Q^2$ for comparison to experiment.

\begin{figure}[tb]
\begin{center}
\begin{minipage}[t]{17.5 cm}
\epsfig{file=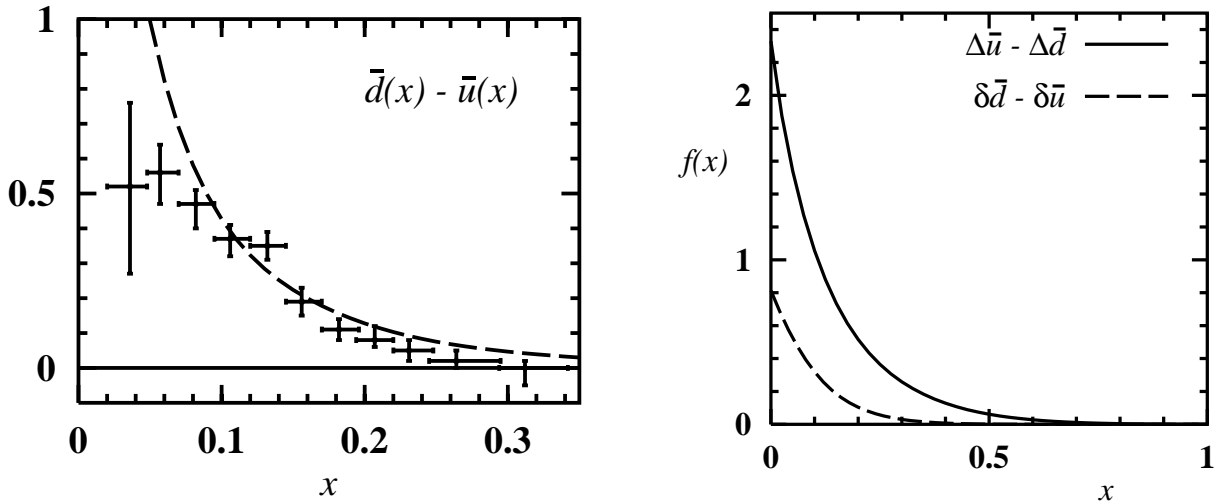,scale=1.0}
\end{minipage}
\begin{minipage}[t]{16.5 cm}
\caption{Calculations from a chiral-quark soliton model~\cite{dres98}
(a) The value 
calculated for $\bar d(x) - \bar u(x)$ in the proton compared to early
data from E866. The calculation has been evolved from $Q=0.6$ GeV to $Q = 7.35$
GeV. (b) The calculated flavor asymmetry of the longitudinally 
polarized antiquark
distribution in the proton (solid line) and the transversity distribution of 
the antiquarks (dashed line). Both distributions are at $Q = 0.6$ GeV.}
\label{fig:4.4.2}
\end{minipage}
\end{center}
\end{figure}

Quarks are described by single particle wave functions which are
solutions of the Dirac equation in the field of the background pions. The
spectrum of single particle states includes a single bound state level 
plus distorted positive and negative Dirac continumn. The discrete level 
and the negative continuum are occupied producing a state of unit baryon 
number. The distribution $\Delta \bar u(x) - \Delta \bar d(x)$
appears in leading order ($N_c^2$) in a $1/N_c$
expansion while $\bar u(x) - \bar d(x)$ appears in next-to-leading order. 
For example the isovector spin carried by quarks is
\begin{eqnarray}
\Delta u(x) - \Delta d(x) = - {1 \over 3} (2T_3)N_c M_N \sum_{n(occup)}
\int {d^3k \over (2\pi)^3} \tilde \Psi^*_n ({\bf k}) (1+\gamma^0 \gamma^3)
\gamma_5\tau^3
\delta(k_3 + E_n +xM_n) \Psi_n({\bf k}).
\label{eq:4.7.1}
\end{eqnarray}
Where $\Psi_n({\bf k})$ are the single particle wave functions, and 
$2T_3 = \pm 1$ for the proton and neutron respectively. The sum runs 
over all occupied single particle quark states, the bound state and 
the negative continuum. Figure~\ref{fig:4.4.2} shows 
some of the results from Ref.~\cite{dres98}.

The values for the integral of the quantities shown in Fig.~\ref{fig:4.4.2}
are
\begin{eqnarray}
\int_0^1 [\bar d(x) - \bar u(x)] dx = 0.17;~~~~~~~
\int_0^1 [\Delta \bar d(x) - \Delta \bar u(x)] dx = -0.31.
\label{eq:4.7.2}
\end{eqnarray}
Equation~\ref{eq:4.7.2} presents a value for the integrated flavor 
asymmetry that is larger, but in the same ballpark as the measured 
value ($0.118 \pm 0.012$). The distribution in $x$ is not unlike 
that of models employing virtual mesons and hence should have a 
similar spatial distribution.  However, the magnitude of the integrated 
isovector longitudinal-spin distribution for the antiquarks
is surprisingly large and totally at variance with the results of models
employing GBs to account for the flavor asymmetry. The GB models obviously
require $\Delta \bar d(x) = \Delta \bar u(x) = 0$. While 
one might think that the 
chiral soliton model should be consistent with the GB models as the 
baryonic soliton is developed in a background field of pions, the 
result for the spin carried by the antiquarks is totally at odds with 
any GB description. Hence one cannot view this non-pertubative approach 
as generating pseudoscalar mesons. Recent data~\cite{smc98} showed 
$\Delta \bar q = 0.01 \pm 0.04 \pm 0.03$, consistent with a small 
sea polarization. We await 
the judgement of future experiments to decide which description best conforms 
to reality.

\section{Further Implications of the Meson-Cloud Models}
\subsection{\it Strange Sea of the Nucleon}

Models in which virtual mesons are admitted as degrees of freedom
have implications that extend beyond the $\bar d, \bar u$ 
flavor asymmetry addressed above.
They create hidden strangeness in the nucleon via such virtual processes as 
$p \to \Lambda + K^+, \Sigma + K$, etc. 
Such processes are of considerable interest as they imply different $s$ and
$\bar s$ parton distributions in the nucleon, a feature not found in gluonic
production of $s \bar s$ pairs. This subject has 
received a fair amount of attention
in the literature~\cite{holt96,signal87,warr92,ji95,ma96} 
but experiments have yet to clearly identify such a
difference. Thus in contrast to the $\bar d, \bar u$
flavor asymmetry, to date there is no positive experimental evidence
for $s \bar s$ contributions to the nucleon from virtual 
meson-baryon states~\cite{conrad98,ccfr95}.

A difference between the $s$ and $\bar s$ distribution can be made manifest
by direct measurement of the $s$ and $\bar s$ parton distribution functions 
in DIS neutrino scattering, or in
the measurement of the $q^2$ dependence of the 
strange quark contribution ($F^p_{1s}(q^2)$)
to the proton charge form factor. 
This latter case is not well known and follows from a suggestion of Kaplan 
and Manohar~\cite{kaplan88} regarding the new information contained in the
weak neutral current form factors of the nucleon. Measurement of these form
factors allows extraction of the strangeness contribution to the 
nucleon's charge and magnetic moment and axial form factors. The portion 
of the charge form factor $F^p_{1s} (q^2)$ due to strangeness clearly is
zero at $q^2 = 0$, but if the $s$ and $\bar s$ distributions are different the
form factor becomes non-zero at finite $q^2$. These ``strange'' form 
factors can be measured in neutrino elastic scattering~\cite{garvey93}
from the nucleon, or by selecting the parity-violating component of
electron-nucleon elastic scattering, as is now
being done at the Bates~\cite{mueller} and Jefferson Laboratories~\cite{aniol}.

It is worth pointing out that there is a relationship between the parton
distributions and the form factors of a hadron.  If the neutron's charge 
form factor is explained in terms of a particular meson-baryon expansion,
then one should expect that the expansion is consistant with the neutron''s
partonic structure. Little work appears to have been done bringing these
descriptions together. Below is an example showing the impact of the
nucleon's pionic content, inferred from the flavor asymmetry, on the spin
distribution in the nucleon. 

\subsection{\it Sea Quark Distributions in Hyperons}

Dilepton production using meson or hyperon beams offers a
means of determining parton distributions of these unstable hadrons. 
Many important features of nucleon parton distributions, such as
the flavor structure and the nature of the non-perturbative sea, find their
counterparts in mesons and hyperons. Information about meson and hyperon 
parton structure could provide valuable new insight into 
nucleon parton distributions. Furthermore,
certain aspects of the nucleon structure, such as the strange 
quark content of the nucleon, could be 
probed with kaon beams.

No data exist for hyperon-induced dilepton production. The observation
of a large $\bar d / \bar u$ asymmetry in the proton has 
motivated Alberg et al.~\cite{alberg1,alberg2}
to consider the sea-quark distributions in the $\Sigma$. The meson-cloud
model implies a $\bar d / \bar u$ asymmetry in the $\Sigma^+$ even larger than
that of the proton. However, the opposite effect is expected
from SU(3) symmetry.
Although relatively intense
$\Sigma^+$ beams have been produced for recent experiments at Fermilab,
this experiment appears to be very challenging because of
large pion, kaon, and proton contaminations in the beam.

\begin{figure}[tb]
\begin{center}
\begin{minipage}[t]{11 cm}
\epsfig{file=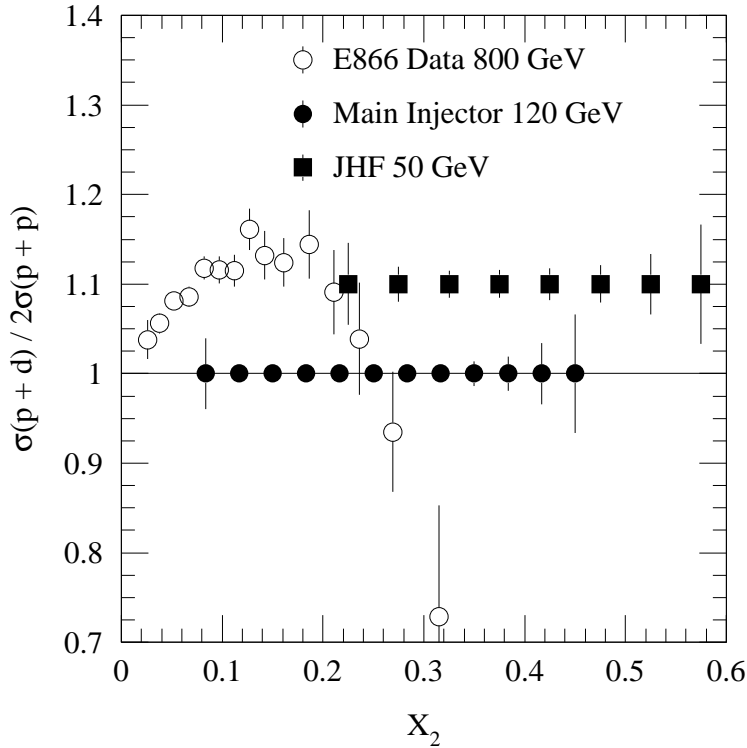,scale=1.2}
\end{minipage}
\begin{minipage}[t]{16.5 cm}
\caption{Projected statistical accuracy for $\sigma (p+d) /2\sigma (p+p)$ in a
100-day run at JHF~\cite{brown,peng00}. The E866 
data and the projected sensitivity for
a proposed measurement~\cite{p906} at the 120 GeV
Fermilab Main-Injector are also shown.}
\label{fig:6.1.1}
\end{minipage}
\end{center}
\end{figure}

\section{Future Prospects}
\subsection{\it $\bar d / \bar u$ at Large and Small $x$}

The interplay between the perturbative and non-perturbative components of
the nucleon sea remains to be better determined. Since the perturbative
process gives a symmetric $\bar d/ \bar u$ while a non-perturbative 
process is needed to generate an asymmetric $\bar d/ \bar u$ sea, the relative
importance of these two components is directly reflected in the $\bar d/ \bar u$
ratios. Thus, it would be very important to extend the DY
measurements to kinematic regimes beyond the current limits. 

The new 120 GeV Fermilab Main Injector (FMI) and the proposed 50 GeV 
Japanese Hadron Facility~\cite{nagamiya} (JHF) present opportunities for 
extending the $\bar d/ \bar u$ measurement to larger $x$ ($x > 0.25$).
For given values of $x_1$ and $x_2$ the DY cross section
is proportional to $1/s$, hence the DY cross section 
at 50 GeV is roughly 16 times greater than
that at 800 GeV! Figure~\ref{fig:6.1.1} shows the expected statistical 
accuracy for $\sigma (p+d)/ 2 \sigma (p+p)$ at JHF~\cite{brown,peng00} compared 
with the data from E866 and a proposed measurement~\cite{p906} using 
the 120 GeV proton beam at the FMI. A definitive measurement of
the $\bar d/ \bar u$ over the region $0.25 < x < 0.7$ could indeed be
obtained at FMI and JHF.

\begin{figure}[tb]
\begin{center}
\begin{minipage}[t]{11 cm}
\epsfig{file=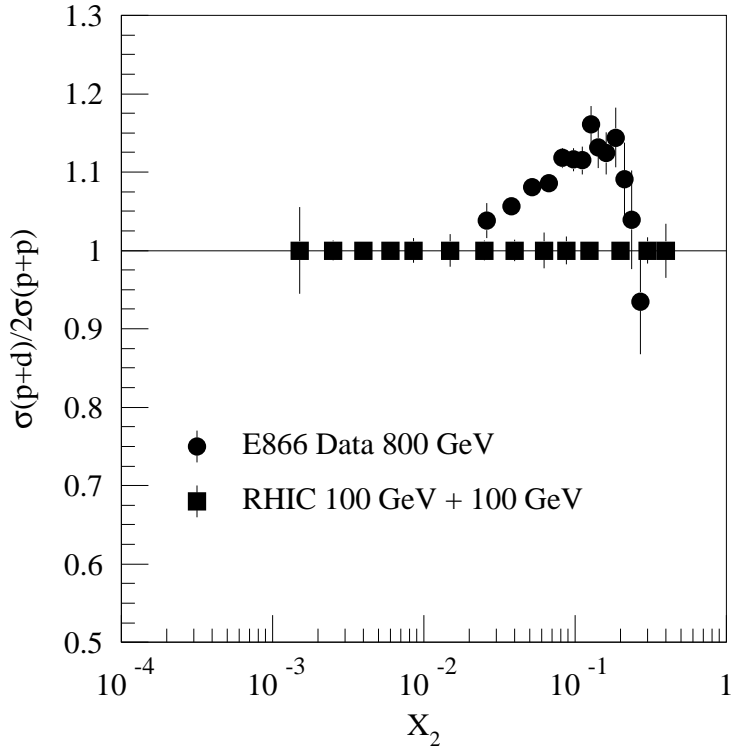,scale=1.2}
\end{minipage}
\begin{minipage}[t]{16.5 cm}
\caption{Projected statistical accuracy for $\sigma (p+d) /2\sigma (p+p)$ in a
two-month run at RHIC using the PHENIX detector~\cite{peng00a}.
The E866 
data are also shown.}
\label{fig:6.1.1.1}
\end{minipage}
\end{center}
\end{figure}

At the other end of the energy scale, RHIC will operate soon in the range
$50 \le \sqrt s \le 500$ GeV/nucleon. The capability of accelerating and
colliding a variety of beams from $p + p$, $p + A$, to $A + A$ at RHIC
offers a unique opportunity to extend the DY $\bar d / \bar u$ measurement
to very small $x$. Such information is
important for an accurate determination of the integral of $\bar d - \bar u$,
as well as for a better understanding of the origins for flavor asymmetry.
The statistical accuracy for measuring $\sigma(p+d)/2\sigma(p+p)$ in a
two-month PHENIX run is shown in Fig.~\ref{fig:6.1.1.1}. Also shown in
Fig.~\ref{fig:6.1.1.1} are the data from E866.
The lowest $x_2$ reachable at
RHIC is around $10^{-3}$, an order of magnitude lower than in E866.

\subsection{\it W Production}

\begin{figure}[tb]
\begin{center}
\begin{minipage}[t]{11 cm}
\epsfig{file=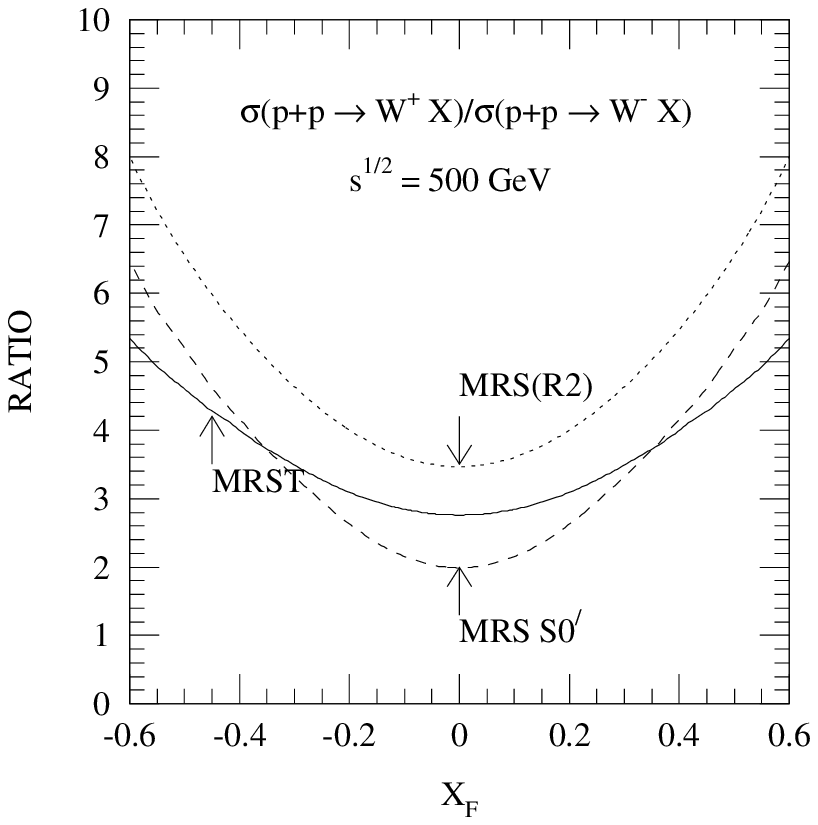,scale=1.2}
\end{minipage}
\begin{minipage}[t]{16.5 cm}
\caption{ Predictions of
$\sigma (p+p \to W^+ X) / \sigma (p+p \to W^- X)$ as a function of $x_F$
at $\sqrt s$ = 500 GeV.
The dashed curve corresponds to the $\bar
d/\bar u$ symmetric MRS S0$^\prime$ structure
functions, while the solid and dotted curves
are for the $\bar d/\bar u$ asymmetric structure function MRST and MRS(R2),
respectively.}
\label{fig:6.2.1}
\end{minipage}
\end{center}
\end{figure}

To disentangle the $\bar d / \bar u$ asymmetry from the possible 
charge-symmetry violation effect~\cite{ma1,londergan2,blt98}, 
one could consider $W$ boson production, a generalized
DY process, in $p + p$ collision at RHIC.
An interesting quantity to be measured is the ratio of the 
$p + p \to W^+ + X$ and $p + p \to W^- + X$ cross sections~\cite{peng1}. 
It can be shown that this
ratio is very sensitive to $\bar d / \bar u$. An important feature of
the $W$ production asymmetry in $p + p$ collision is that it is completely free 
from the assumption of charge symmetry. Figure~\ref{fig:6.2.1} shows the 
predictions for $p + p$ collision at $\sqrt s =
500~$GeV. The dashed curve corresponds to the $\bar
d/\bar u$ symmetric MRS S0$^\prime$~\cite{mrss0} structure 
functions, while the solid and dotted curves
are for the $\bar d/\bar u$ asymmetric structure function MRST and MRS(R2),
respectively. Figure~\ref{fig:6.2.1} clearly shows that $W$ asymmetry 
measurements at RHIC could provide an independent determination 
of $\bar d / \bar u$.

\subsection{\it Strange Sea in the Nucleon}

As discussed earlier, an interesting consequence of the meson-cloud model
is that the $s$ and $\bar s$ distributions
in the proton could have very different shapes, even though the net amount
of strangeness in the proton vanishes. By 
comparing the $\nu$ and $\bar \nu$
induced charm production, the CCFR 
collaboration found no difference between 
the $s$ and $\bar s$ distributions~\cite{ccfr95}. More precise
future measurements would be very helpful.
Dimuon production experiments using $K^\pm$ beams might provide an independent
determination of the $s$/$\bar s$ ratio of the proton, provided that our
current knowledge on valence-quark distributions in kaons is improved.
As discussed in Section 5.1, ongoing measurements of $F^p_{1s}$ via 
parity-violating electron-nucleon scattering should shed much light on
the possible difference between $s$ and $\bar s$ distributions.

\subsection{\it Sea Quark Polarization}

Polarized DY and $W^\pm$ production in polarized $p+p$ collision 
are planned at RHIC~\cite{bunce} and they have great potential for providing
qualitatively new information about antiquark polarization. At large
$x_F$ region ($x_F > 0.2$), the longitudinal spin asymmetry $A_{LL}$ in the
$p+p$ DY process is given by~\cite{moss,plm}
\begin{eqnarray}
A^{DY}_{LL}(x_1,x_2) \approx g_1(x_1)/F_1(x_1) \times {\Delta \bar u
\over \bar u}(x_2),
\label{eq:6.4.1}
\end{eqnarray}  
where $g_1(x)$ is the proton polarized structure function measured in DIS,
and $\Delta \bar u(x)$ is the polarized $\bar u$ distribution function.

Equation~\ref{eq:6.4.1} shows that $\bar u$ polarization can be determined using 
polarized DY at RHIC. Additional information on the sea-quark polarization
can be obtained via $W^\pm$ production~\cite{bs93}.
The parity-violating
nature of $W$ production implies that only one of the two beams need to
be polarized. At positive $x_F$ (along the direction of the polarized
beam), one finds\cite{bs93},
\begin{eqnarray}
A_L^{W^+}\approx {\Delta u \over u}(x_2),\ \ {\rm and}\ \ \
A_L^{W^-}\approx {\Delta d\over d}(x_2),  \label{eq:6.4.2} 
\end{eqnarray}
where $A^W_L$ is the single-spin asymmetry for $W$ production.
Equation~\ref{eq:6.4.2} shows that the flavor dependence of the sea-quark
polarization can be revealed via $W^\pm$ production at RHIC.
A remarkable prediction of the chiral quark-soliton model is that
the flavor asymmetry of polarized sea-quark, $\Delta \bar u(x) -
\Delta \bar d(x)$, is very large~\cite{dres99a}. This is in striking
contrast to the meson cloud model which predicts very small 
values for $\Delta \bar u(x) - \Delta \bar d(x)$~\cite{fries98,boreskov99}.
Future DY and $W^\pm$ production experiments at RHIC could clearly test
these models~\cite{dres99b}

\subsection{\it Hard Processes Tagged by Leading Particles}

The role of the meson-cloud model in explaining the $\bar d / \bar u$
asymmetry suggests a novel technique to study meson substructures
without using a meson beam. The idea is that the meson cloud in the nucleon    
could be considered as a virtual target to be probed by various hard       
processes. Recently at the HERA e-p collider, meson structure functions
were measured in a hard diffractive process, where forward-going neutrons
or protons were tagged in coincidence with the DIS events~\cite{h1}.
Analogous measurements could be done at RHIC for p-p collisions~\cite{peng00a}.
In particular, a DY pair in coincidence with a forward-going neutron
or proton could provide information on the antiquark distributions in pions
at small $x$. The underlying picture of the tagged DY process is 
as follows. A proton fluctuates into a baryon plus a meson 
( $p \to n + \pi^+$, $p \to \Lambda + K^+$, for
example) and the other proton beam interacts with the meson,
producing a lepton pair. The remaining baryon moves roughly
along the initial proton beam direction and could be detected at
forward angles.
A simulation for the $p + p \to n + \mu^+ \mu^- + X$ for the PHENIX detector,
where the muon pairs are detected in the muon arms and the neutrons
are detected by a small-angle calorimeter showed that such measurement
is quite feasible~\cite{peng00a}.

Drell-Yan experiment tagged by forward-going baryons at RHIC would provide
a direct test of the meson-cloud model. One could also conceive tagging
forward-going $\Delta$ or $\Lambda$. The $\Lambda$-tagging is of special
interest since it can shed light on the strange-quark contents of the
proton. Another extension is the measurement of double-helicity asymmetry,
$A_{LL}$, in neutron-tagged $\vec p + \vec p$ Drell-Yan process. If
the dominant underlying process is $\vec p + \pi$ interaction, $A_{LL}$
is expected to be zero. 

\section{Conclusion}

The flavor asymmetry of the nucleon sea has been clearly established
by recent DIS and DY experiments. The surprisingly large asymmetry
between $\bar u$ and $\bar d$ is unexplained by perturbative
QCD. We draw the following conclusions:
\begin{itemize}

\item The $x$ dependence of $\bar d_p / \bar u_p$ indicates that a
$\bar d, \bar u$ symmetric sea dominates at small ($x < 0.05$) and 
large $x$ ($x > 0.3$). But for $0.1 < x < 0.2$ a large and significant
flavor non-symmetric contribution determines the sea distributions.

\item The value of the integrals
\begin{eqnarray}
\int_0^1 [\bar d_p(x,Q^2) - \bar u_p (x,Q^2)] dx
= \int_0^1 [\bar u_n(x,Q^2) - \bar d_n (x,Q^2)] dx = 0.118 \pm 0.012
\label{eq:7.1}
\end{eqnarray}
are approximately independent of $Q^2$ and are characteristic of the nucleon.

\item The value of the integral and the distribution in $x$ can be
reproduced using virtual pions. The $x$ distribution favors the virtual pion
being emitted by the nucleon as a whole, rather than by a constituent quark.

\item Virtual pion emission reduces the spin carried by quarks but is
not sufficient to account for the observed reduction.

\item The mean square charge in the neutron sea is less than that in the
proton sea, because $\bar u_n > \bar d_n$ and $\bar d_p > \bar u_p$.
\end{itemize}

The up-down flavor asymmetry in the nucleon sea is as fundamental as
the axial vector coupling constant, $g_A$, in characterizing the
properties of the nucleon. This asymmetry is of a
completely non-perturbative origin but, differently from $g_A$, could
only be observed by measurements at the partonic level. Thus it stands as
a key property of the nucleon that awaited the discovery of perturbative QCD
processes and the associated high-energy facilities
to uncover it. While the nucleon ``spin crisis" involves both
perturbative (spin carried by gluons) and non-perturbative processes
(spin flips due to the emission of virtual GBs),
the origin of the up-down asymmetry in the nucleon sea is entirely
non-perturbative. Fortunately, it is most likely 
that the origin of this asymmetry 
lies with the presence of virtual isovector mesons, mostly pions, in
the nucleon sea. The unfortunate aspect of this problem is that 70 years
of research in nuclear physics and the structure of the nucleon is still
unable to provide a quantitative answer to this basic question.
Probably a clear picture would eventually emerge after some future
experimental investigation which includes:

\begin{itemize}

\item Measurement of $\bar d_p/ \bar u_p$ for $x > 0.25$ and $x < 0.01$ 
to more firmly establish the origin of the flavor asymmetry.

\item $\Delta \bar d$ and $\Delta \bar u$ in the interval $0.1 < x < 0.2$ 
needs to be measured. Each must be $\sim 0$ if the pionic explanation of 
the $\bar d, \bar u$ asymmetry is correct.
The chiral soliton model predicts that $\Delta \bar d - \Delta \bar u \sim
-0.5$ in the proton for this $x$ interval.

\item Semi-inclusive including polarized  DIS and Drell-Yan experiments
to demonstrate the role of mesonic degrees of freedom in the nucleon's
makeup.

\item More precise measurements on the $s$ versus $\bar s$ distributions in
the nucleon.

\item Experimental identification of instantons.

\end{itemize}

\vskip 0.4in
\noindent{Acknowledgment}

We are grateful to our collaborators on the Fermilab E772 and E866.
This work 
was supported by the US Department of Energy, Nuclear Science 
Division, under contract W-7405-ENG-36.

\end{document}